\documentclass[10pt,twocolumn,letterpaper]{article}

\usepackage{cvpr}
\usepackage{times}
\usepackage{epsfig}
\usepackage{graphicx}
\usepackage{amsmath}
\usepackage{amssymb}
\usepackage{color}
\usepackage{times}
\usepackage{epsfig}
\usepackage{algorithm}
\usepackage{algorithmicx}
\usepackage{algpseudocode}
\usepackage{graphics}
\usepackage{threeparttable}
\usepackage{color}
\usepackage[normalem]{ulem}
\usepackage{multirow}
\usepackage{float}
\usepackage{amsfonts}
\usepackage{bm}
\usepackage{subfig}
\usepackage{array}
\usepackage[table]{xcolor}
\usepackage{colortbl}
\usepackage{pifont}
\usepackage{cite}
\usepackage{diagbox}
\usepackage{rotating}
\usepackage{booktabs}
\usepackage{overpic}
\usepackage{textcomp}
\usepackage{contour}
\usepackage{enumitem}

\usepackage[pagebackref=true,breaklinks=true,letterpaper=true,colorlinks,bookmarks=false]{hyperref}
\cvprfinalcopy 

\def\cvprPaperID{****} 

\ifcvprfinal\fi
\begin{document}

\title{Bridge the Vision Gap from Field to Command: A Deep Learning Network Enhancing Illumination and Details}

\author{Zhuqing Jiang$^1$$^,$$^2$$^,$\footnotemark[1]\quad 
	Chang Liu$^{1,}$\footnotemark[1] \footnotemark[2]\quad    Ya'nan Wang$^1$\quad  \\
	\quad Kai Li$^1$\quad  Aidong Men$^1$\quad Haiying Wang$^1$\quad Haiyong Luo$^3$\\
	\small $^1$Beijing University of Posts and Telecommunications \quad \\
	\small $^2$Beijing Key Laboratory of Network System and Network Culture \quad\\
	\small $^3$Institute of Computing Technology, Chinese Academy of Sciences \quad\\
	{\tt\small $\{$jiangzhuqing, chang\_liu, wynn, xiaoyao125656, menad, why$\}$@bupt.edu.cn \quad yhluo@ict.ac.cn}
}

\maketitle
\thispagestyle{empty}
\renewcommand{\thefootnote}{\fnsymbol{footnote}}
\footnotetext[1]{The first two authors contribute equally to this work.}
\footnotetext[2]{Chang Liu (chang\_liu@bupt.edu.cn) is the corresponding author.}

\vspace{-8pt}
\begin{abstract}
With the goal of tuning up the brightness, low-light image enhancement enjoys numerous applications, such as surveillance, remote sensing and computational photography. Images captured under low-light conditions often suffer from poor visibility and blur. Solely brightening the dark regions will inevitably amplify the blur, thus may lead to detail loss. In this paper, we propose a simple yet effective two-stream framework named NEID to tune up the brightness and enhance the details simultaneously without introducing many computational costs. Precisely, the proposed method consists of three parts: Light Enhancement (LE), Detail Refinement (DR) and Feature Fusing (FF) module, which can aggregate composite features oriented to multiple tasks based on channel attention mechanism. Extensive experiments conducted on several benchmark datasets demonstrate the efficacy of our method and its superiority over state-of-the-art methods.
\end{abstract}

\vspace{-6pt}
\section{Introduction}
\label{sec:intro}

\begin{figure*}[t]
	\centering
	\includegraphics[width=0.85\linewidth]{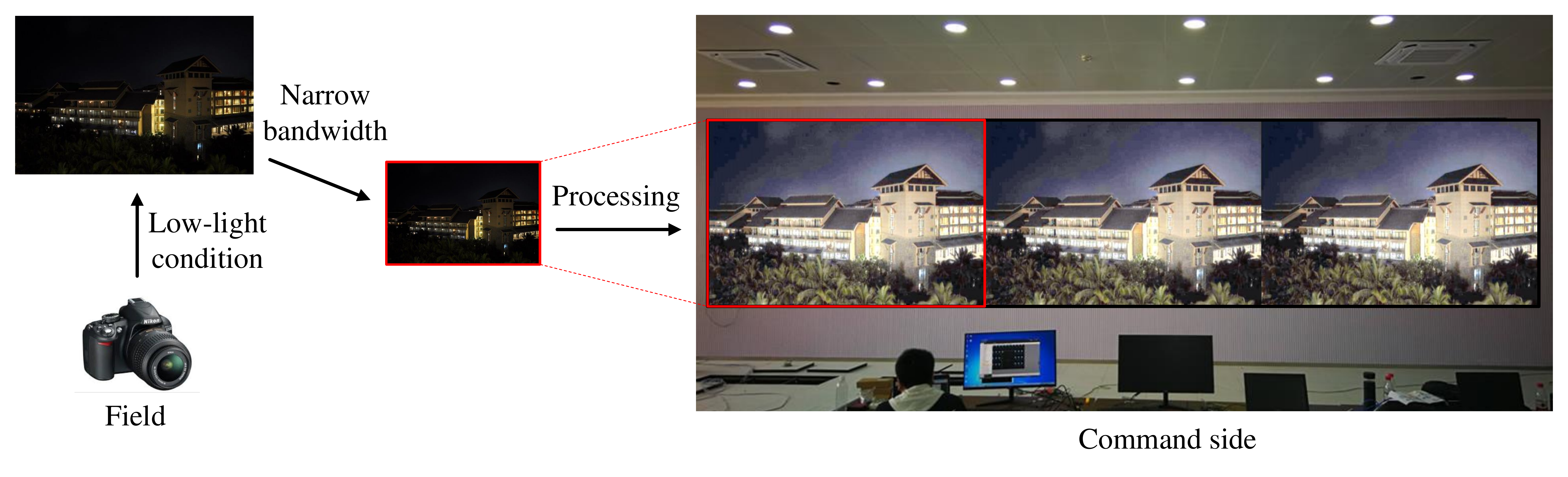}
	\vspace{-0.3cm}
	\caption[]{The process diagram of field-to-command scenarios. In the field, we may take photos under low-light conditions and then transmit images under narrow bandwidth, causing degradation in the image quality. After the processing of our method, the output normal-light and high-resolution images satisfy the requirements of the command side.}
	\label{fig:flowmap}
	\vspace{-0.4cm}
\end{figure*}

In some field tasks such as remote sensing, maritime affairs and environmental protection, we often have to take photos under low-light conditions or transmit images under narrow bandwidth. Thus, the quality of the images is degraded. However, on the command side, bright and detail-rich images are always required, as illustrated in Figure \ref{fig:flowmap}. Therefore, we need a post-processing method that can brighten images and refine details simultaneously.

\begin{figure}[t]
	\centering
	\includegraphics[width=\linewidth]{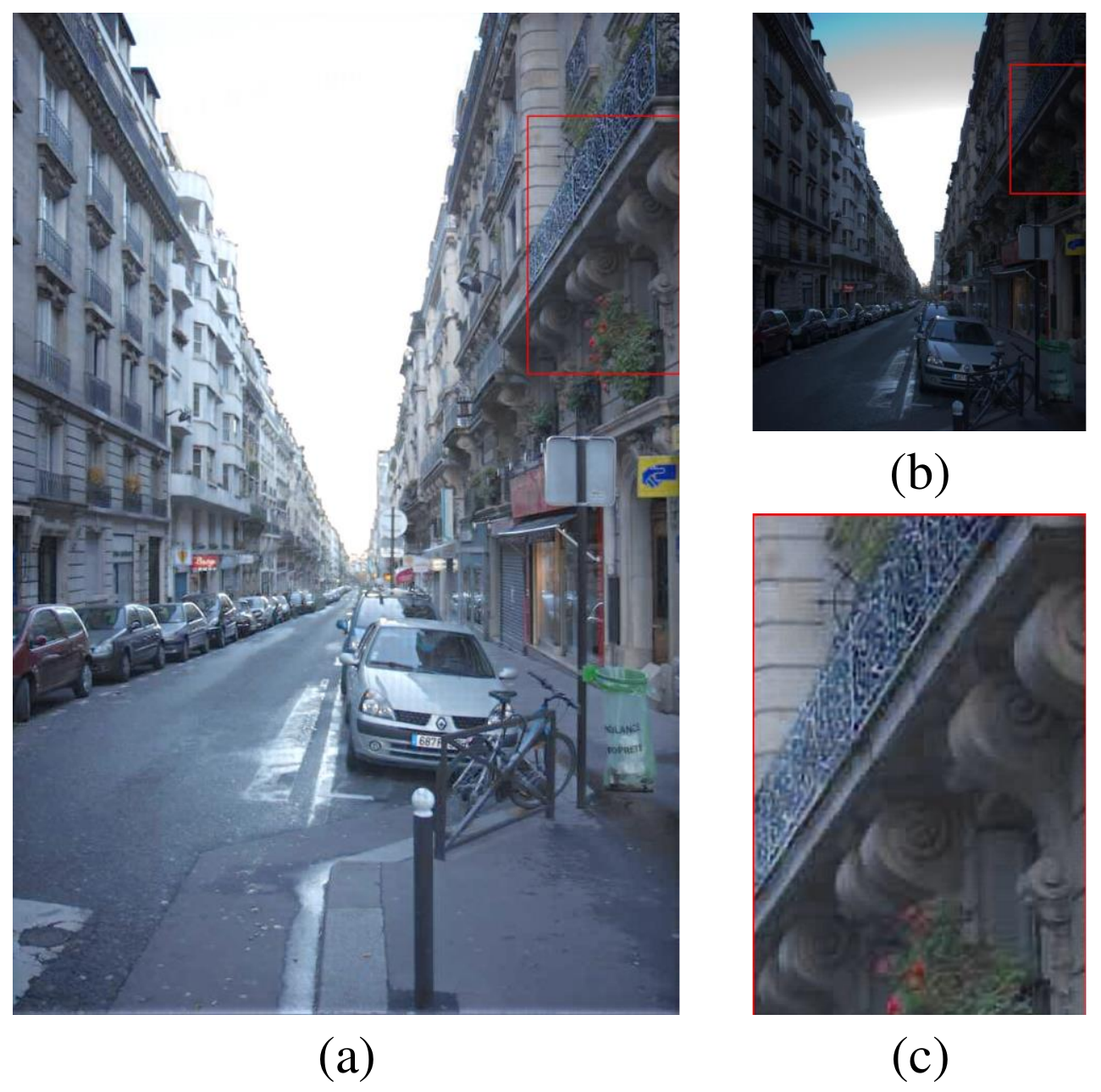}
	\vspace{-0.5cm}
	\caption[]{ (a) is the normal-light and high-resolution version of the low-light image (b) enhanced by the proposed NEID. (c) is a detailed region of (a). Our method brightens up the challenging low-light image while restoring the inherent color and enhancing the details.}
	\label{fig:resultshow}
	\vspace{-0.4cm}
\end{figure}

In the past decades, various methods were proposed to improve the subjective and objective quality of low-light images. Early work \cite{stark2000adaptive,yuan2012automatic,SRIE,wang2017contrast,cai2017joint} primarily focuses on contrast enhancement, which may be insufficient to recover image color and details. With the rapid development of deep neural networks, CNN has been widely used in low-level computer vision tasks, including low-light image enhancement. Recent work \cite{hwang2012context,ignatov2017dslr,gharbi2017deep,chen2018deep,hu2018exposure,park2018distort} takes CNN-based approaches to simultaneously learn adjustment in color, contrast, brightness, and saturation to produce more expressive results. However, these methods still have limitations on brightening severely dark images, causing blurred details.

Therefore, in this paper, we design an effective framework to solve the dilemma. From an example of enhancing a low-light image shown in Figure \ref{fig:resultshow}, the NEID brightens up the image while restoring the inherent color and enhancing the details. Specifically, motivated by image super-resolution, which aims to reconstruct a high-resolution image with a low-resolution input and enhance the details, we propose the novel framework to restore brightness and enhance details simultaneously. Such a learning method is unified in a two-stream framework, consisting of two branches deployed in parallel: Light Enhancement (LE) branch and Detail Refinement (DR) branch. The features of two branches are then fused by a Feature Fusing (FF) module. We integrate the idea of super-resolution into low-light image enhancement pipelines, which utilize a U-net as the codec, thus formulating the LE branch. Then, the encoded features of LE are further enhanced by the fine-grained detail representation from DR with the FF module. It should be noted that these two branches share the same encoder. Finally, DR is optimized with reconstruction supervision only during training, and the post-process of the DR branch will be freely removed from the network in the inference stage, thus causing cost-free overhead.

The main contributions of this work can be summarized as follows:

\noindent 1)
Inspired by image super-resolution, the proposed method integrates low-light image enhancement and image super-resolution, producing normal-light images with rich details and high visual quality.

\noindent 2)
Based on dual codecs, we propose a two-stream network to tackle this unified task, in which the encoders share the same weights.

\noindent 3)
Our proposed network accepts a low-resolution low-light image as the input, which can reduce computational costs and increase inference speed while maintaining or even improving the performance of low-light image enhancement.

\vspace{-5pt}
\section{Methodology}
\label{sec:methodology}

This section first provides a brief review of the most popular U-net\cite{U-net} architecture for low-light image enhancement. We then present the proposed NEID framework in detail and finally introduce the optimization function.

\vspace{-5pt}
\subsection{Review of U-net Framework}
\label{sec:review}

The U-net\cite{U-net} framework is universally applied to low-light image enhancement since it was proposed. This architecture consists of a contracting path and an expansive path. The contracting path consists of the repeated application of two 3$\times$3 convolutions, each followed by a 2$\times$2 max pooling operation with stride 2 for downsampling. The contracting path can be seen as an encoder, which extracts hierarchical features from input low-light images and these features include color, brightness, contrast and saturation information for reconstructing normal-light images. Every step in the expansive path consists of an upsampling of the feature map followed by a 2$\times$2 convolution, a concatenation with the corresponding feature map from the contracting path, and two 3$\times$3 convolutions. The expansive path can be seen as a decoder, which fuses the features and restores the brightness. Moreover, U-net has symmetric skip connections between these two paths, directly propagating low-level features from the encoder to the decoder.

The goal of the low-light image enhancement task is to restore the corresponding normal-light image with the high visual quality given the low-light input image. High visual quality means vivid color and precise details. However, most of the existing U-net architecture methods can merely restore the brightness, while image details are always ignored. 

\begin{figure*}[t]
	\centering
	\includegraphics[width=\linewidth]{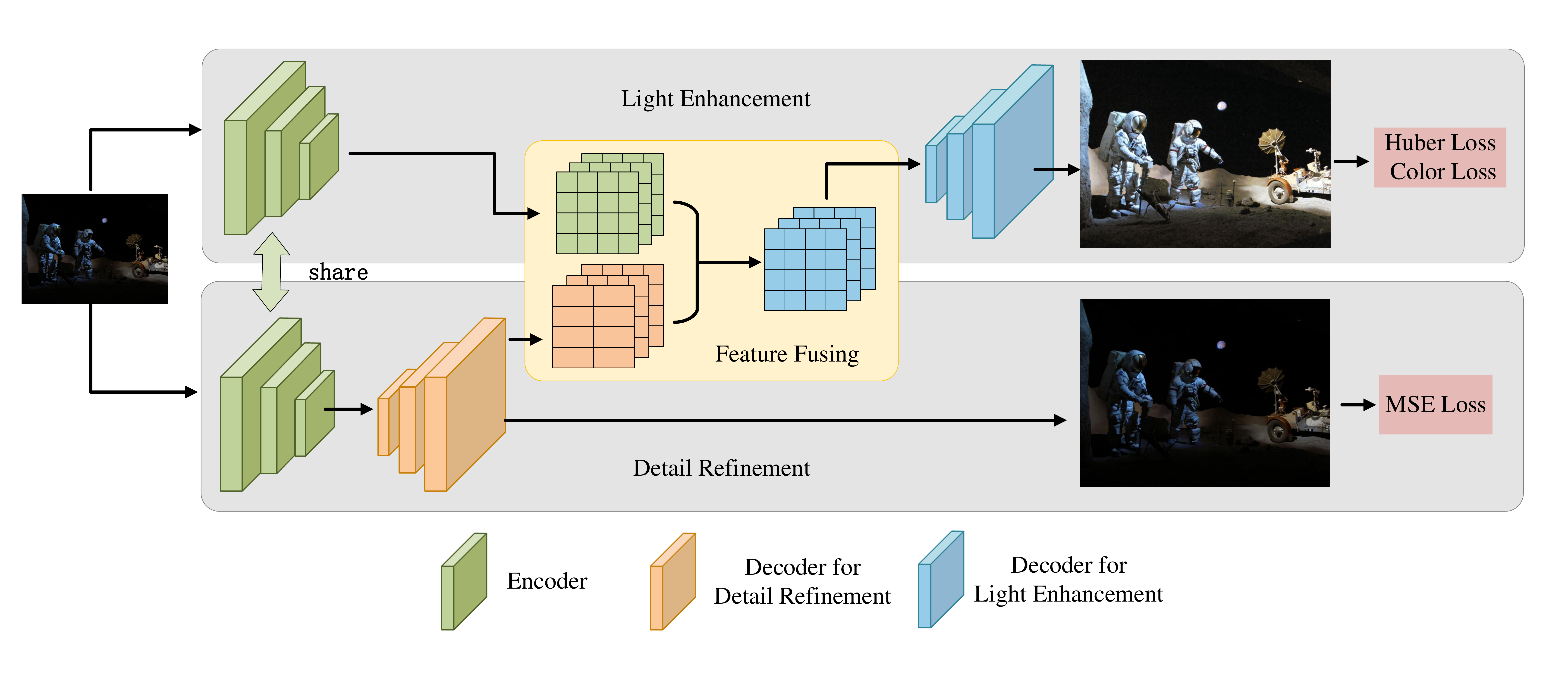}
	\vspace{-0.9cm}
	\caption[]{The overview of the proposed NEID framework, which primarily consists of three parts: Light Enhancement (LE) branch, Detail Refinement (DR) branch, and Feature Fusing (FF) module. The encoder is shared between the LE branch and the DR branch. The overall framework is optimized with three losses: Huber loss for the LE branch, MSE loss for the DR branch and Color loss to relieve color distortion.}
	\label{fig:architecture}
	\vspace{-0.4cm}
\end{figure*}

\vspace{-5pt}
\subsection{Architecture}
\label{sec:architecture}

\noindent\textbf{Consideration \& Motivation}

Low-light image enhancement and super-resolution have to be performed simultaneously under constrained circumstances. Intuitively, an option is to concatenate a low-light image enhancement model and a super-resolution model, propagating normal-light images enhanced by the former model to the latter one for detail refinement. However, this pipeline has serious drawbacks. Low-light images suffer from blurred details besides the extremely low luminance, whose details are barely enhanced by existing low-light image enhancement models. Another issue is that during the training of SR models, input images are generated by downsampling high-definition high-resolution images with bicubic interpolation, which means the blur kernels of these two tasks are different. As a result, super-resolution models cannot restore satisfying high-resolution images facing differently-blurred details.

To alleviate the above dilemmas, we propose a framework to improve the performance without computation and memory overload, especially with a low-resolution input. As shown in Figure \ref{fig:architecture}, our architecture consists of three parts: (a) Light Enhancement (LE); (b) Detail Refinement (DR), and (c) Feature Fusing (FF) module.

\vspace{5pt}
\noindent\textbf{Light Enhancement}

\begin{figure}[t]
	\vspace{0.1cm}
	\centering
	\includegraphics[width=\linewidth]{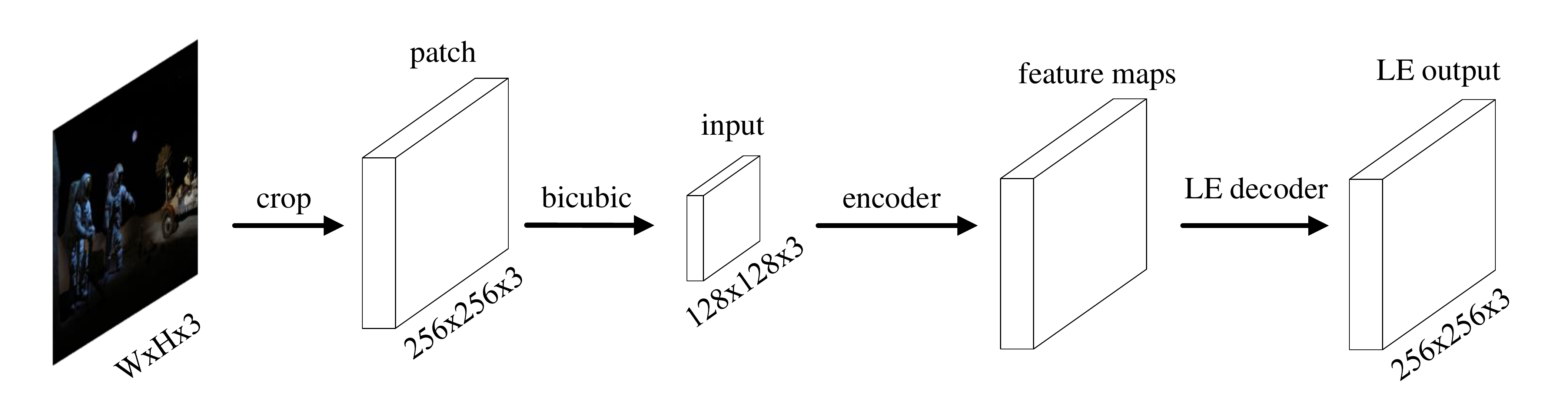}
	\vspace{-0.5cm}
	\caption[]{The LE branch takes an input of 128$\times$128 and generates an output of 256$\times$256.}
	\label{fig:imagesize}
	\vspace{-0.4cm}
\end{figure}

For the LE path, we append an upsampling module after a U-net to generate the final normal-light and high-resolution output. As shown in Figure \ref{fig:imagesize}, during training, LE takes an input of 128$\times$128 and generates an output of 256$\times$256, which is 2$\times$ than the input image. Patch size is known to have a significant impact on low-level image restoration tasks. It is generally acknowledged that a larger patch size results in better model performance yet means large video memory consumption and computational cost. Therefore, a small patch size is chosen to trade off training efficiency against model performance. Our method utilizes a low-resolution input while maintaining the same output size as other methods, which helps accelerate the inference and reduce the computational cost. Compared with other methods that keep the same size of input and output images, our proposed method takes smaller size images as input and extracts features at a smaller resolution, making better use of global information and maintaining the brightness consistency of the entire image. The extra upsampling module applies sub-pixel convolution\cite{VESPCN}.

\begin{table*}[ht]
	\begin{center}
		\vspace{-0.4cm}
		\caption{Quantitative evaluation on the LoL dataset. The highest and second highest results are marked in red and blue.}
		\begin{tabular}{c|cccccccccccccc}
			\toprule[1.5pt]
			& & & & & & & & & & &\\[-8pt]
			\textbf{Method}      & BIMEF  & CRM   & Dong  & LIME  & MF   & Retinex-Net & MSR   & NPE   & GLAD  & KinD  & Ours  \\
			& \cite{BIMEF}   &\cite{CRM}   & \cite{Dong}  & \cite{LIME}  & \cite{MF}    & \cite{Retinex-Net} & \cite{MSR}   & \cite{NPE}   & \cite{GLAD}  & \cite{KinD}   &\\\midrule[1pt]
			& & & & & & & & & & &\\[-8pt]
			\textbf{PSNR$\uparrow$} & 13.88 & 17.20 & 16.72 & 16.76 & 18.79 & 16.77       & 13.17 & 16.97 & 19.72 & \color{blue}20.87   & \color{red}22.83 \\
			& & & & & & & & & & &\\[-8pt]
			\textbf{SSIM$\uparrow$} & 0.58 & 0.64  & 0.58  & 0.56  & 0.64  & 0.56        & 0.48  & 0.59  & 0.70  & \color{red}0.80  & \color{blue}0.78  \\ \bottomrule[1.5pt]
		\end{tabular}
		\label{tab:compare_lol}
	\end{center}
\end{table*}

\begin{figure*}[h]
	\begin{center}
		\begin{tabular}{c@{ }c@{ }c@{ }c@{ }c@{ }}
			\includegraphics[width=0.24\linewidth]{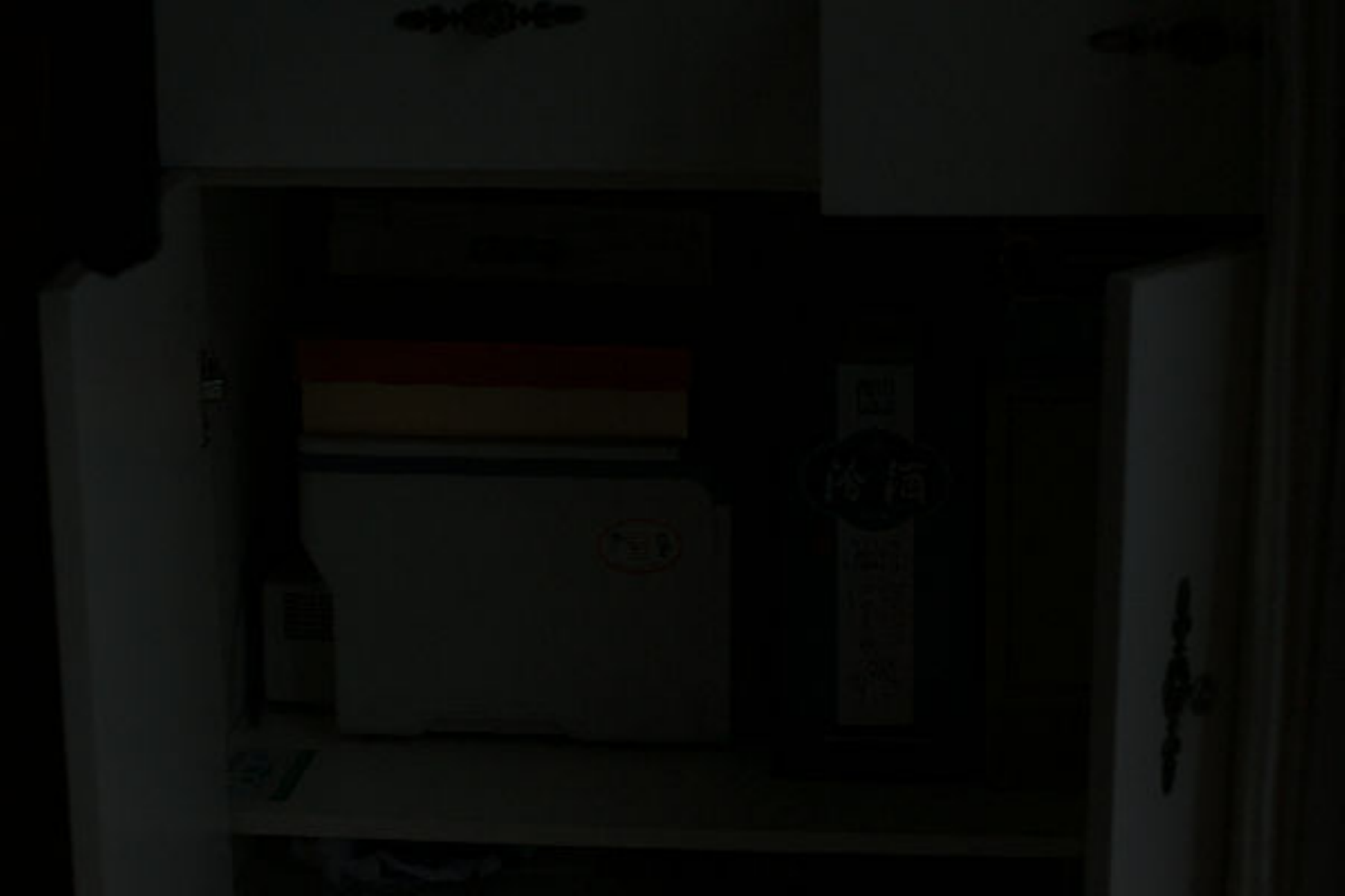}&
			\includegraphics[width=0.24\linewidth]{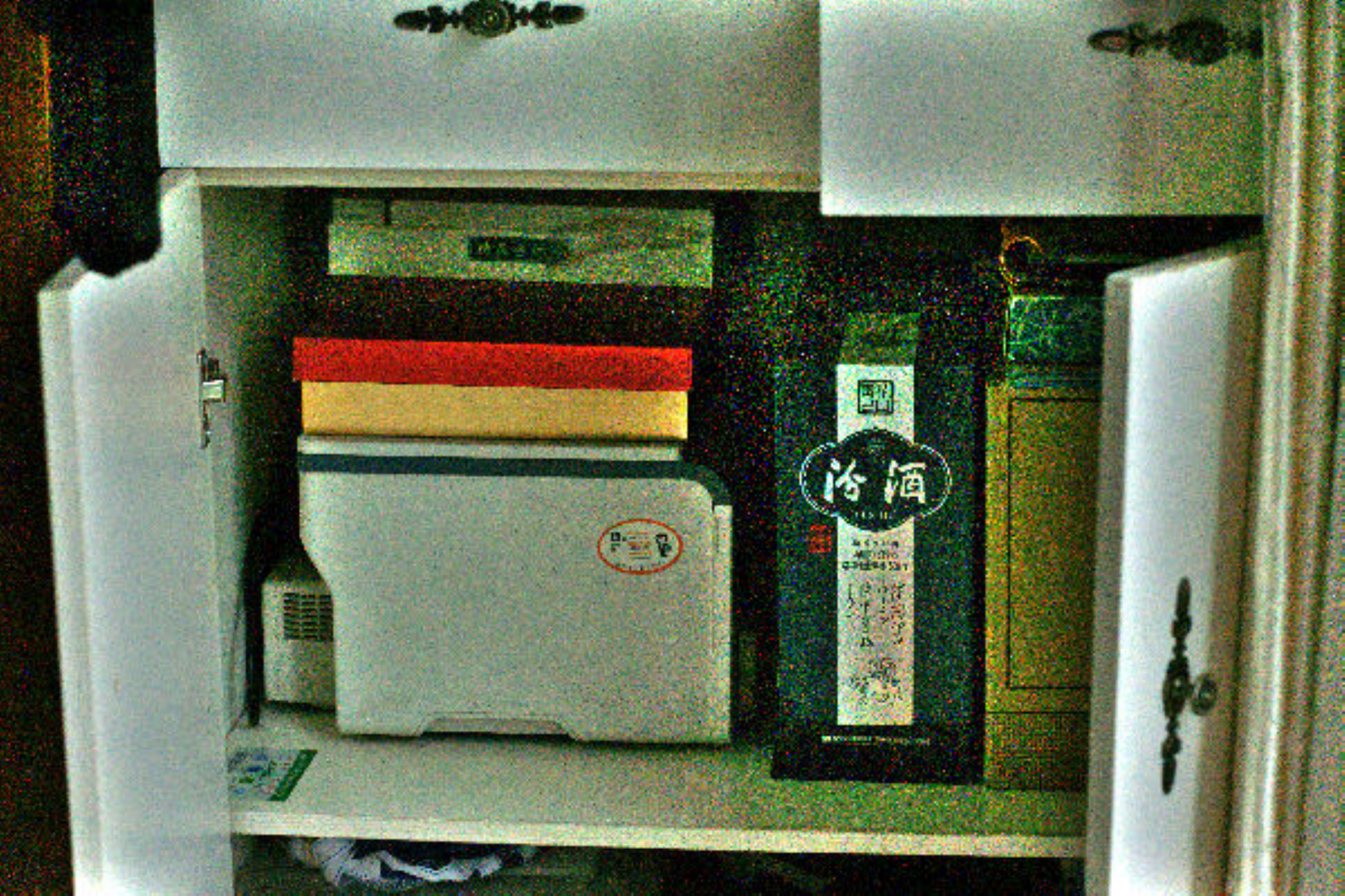}&
			\includegraphics[width=0.24\linewidth]{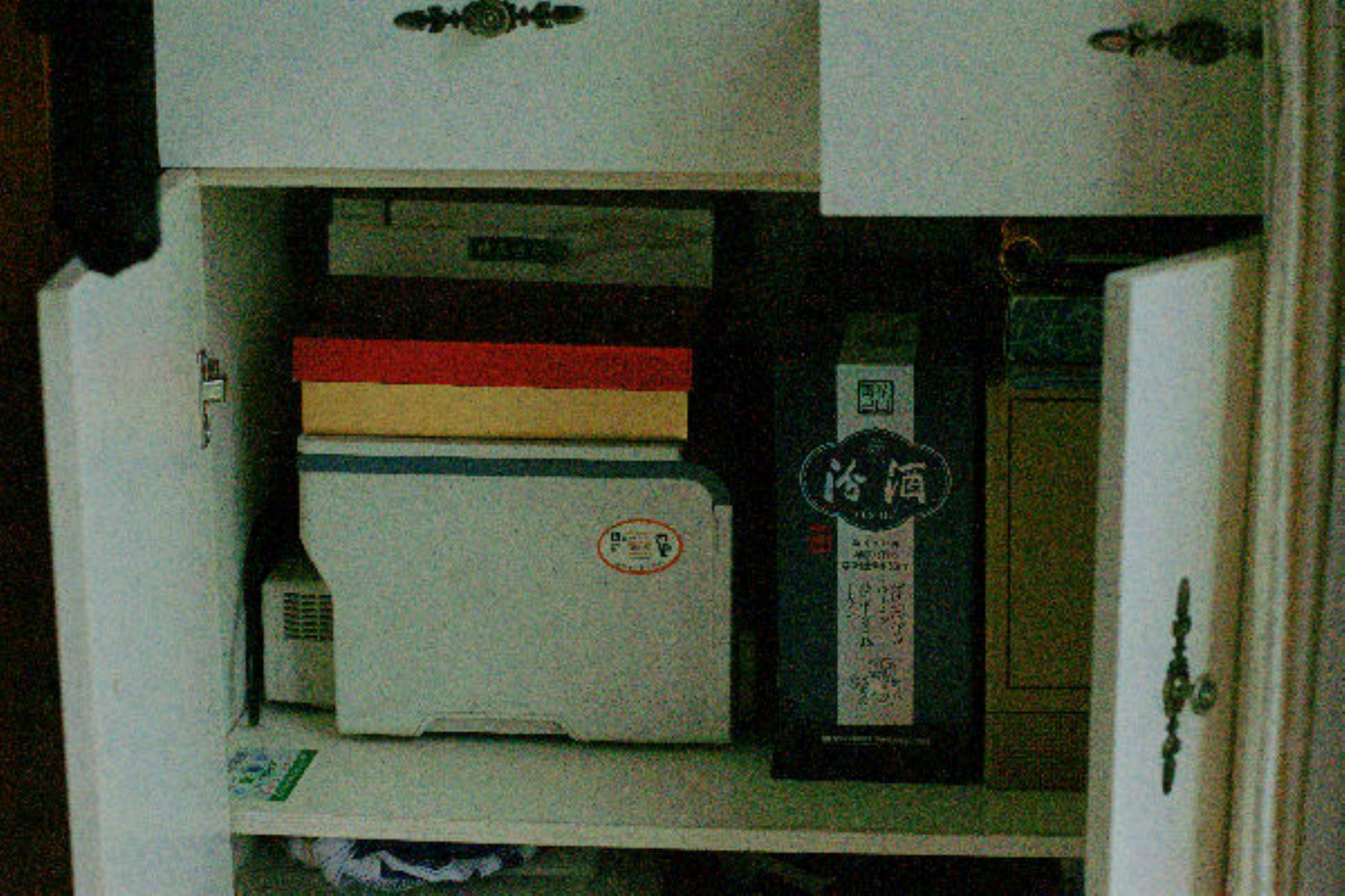}&
			\includegraphics[width=0.24\linewidth]{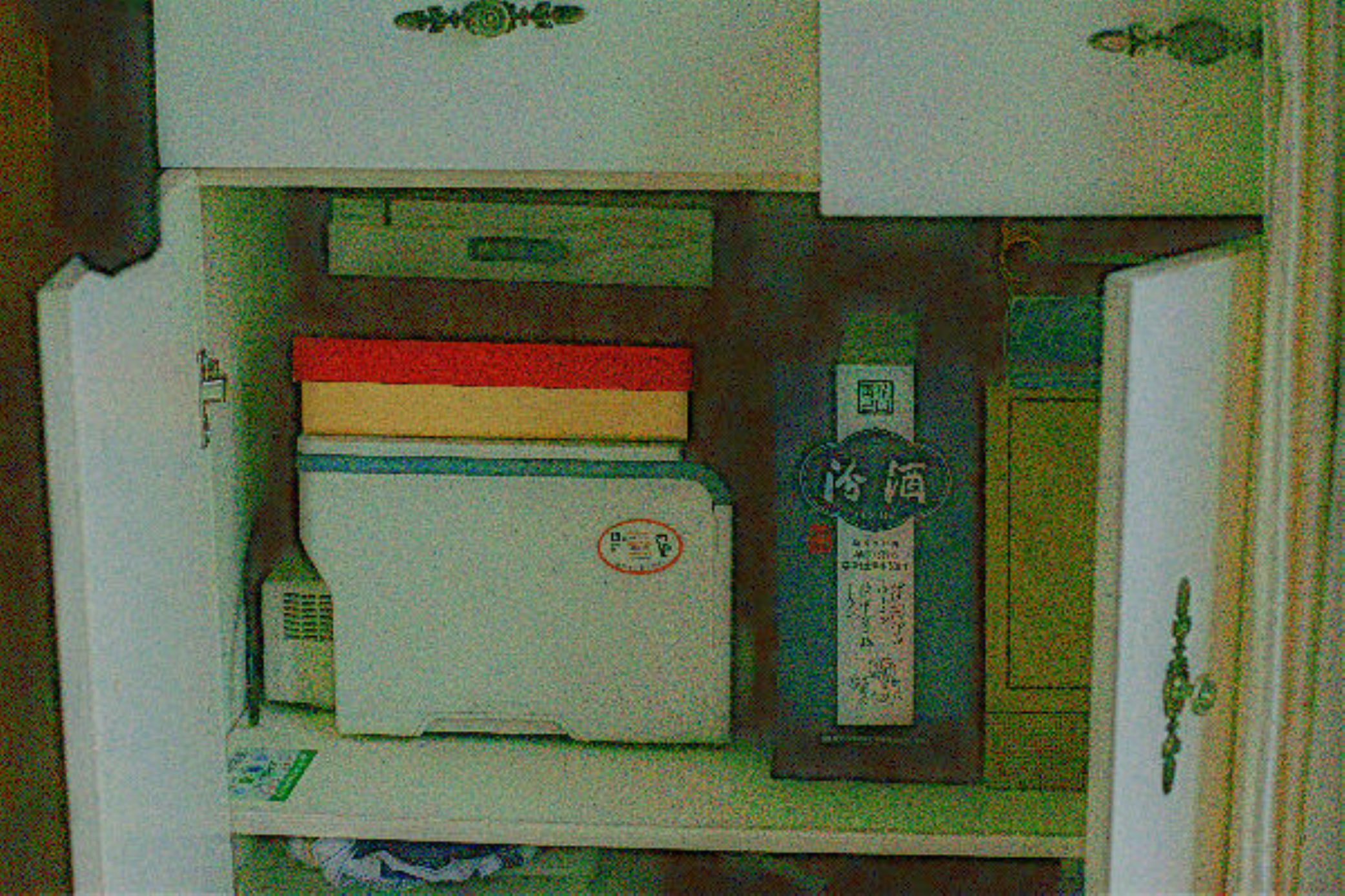}\\
			(a) Input & (b) LIME\cite{LIME} & (c) MF\cite{MF} & (d) Retinex-Net\cite{Retinex-Net} \\
			\includegraphics[width=0.24\linewidth]{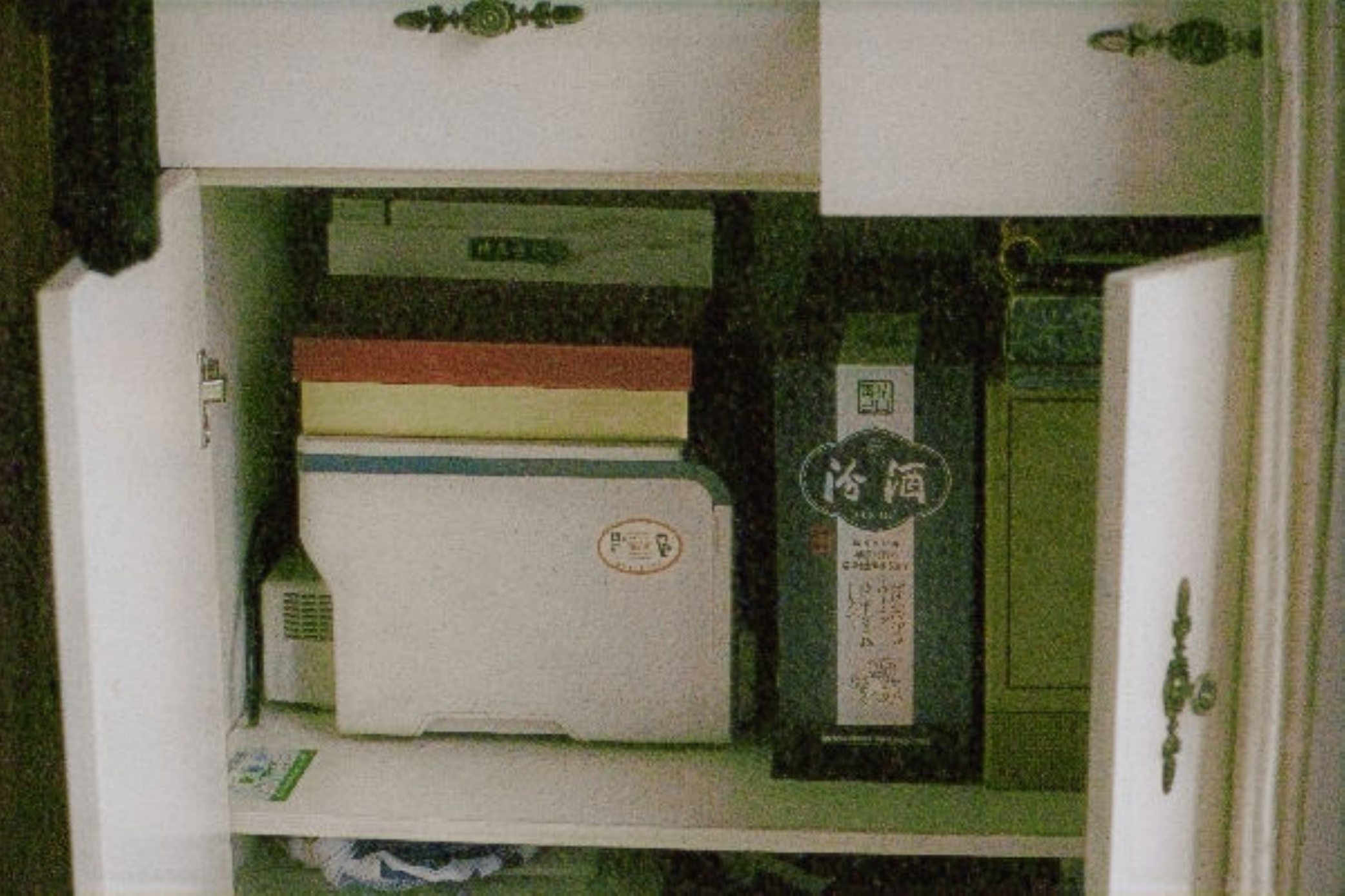}&
			\includegraphics[width=0.24\linewidth]{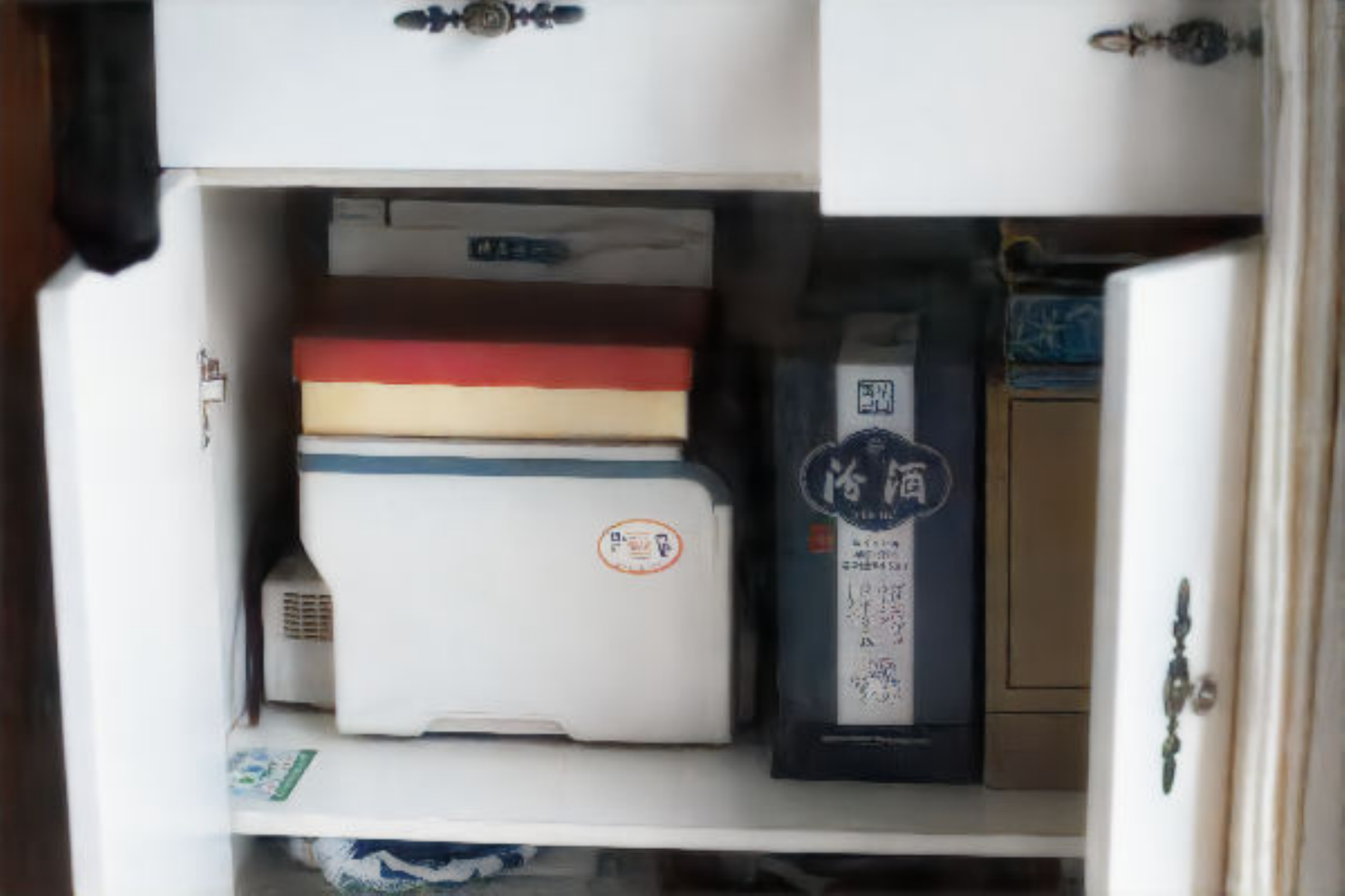}&
			\includegraphics[width=0.24\linewidth]{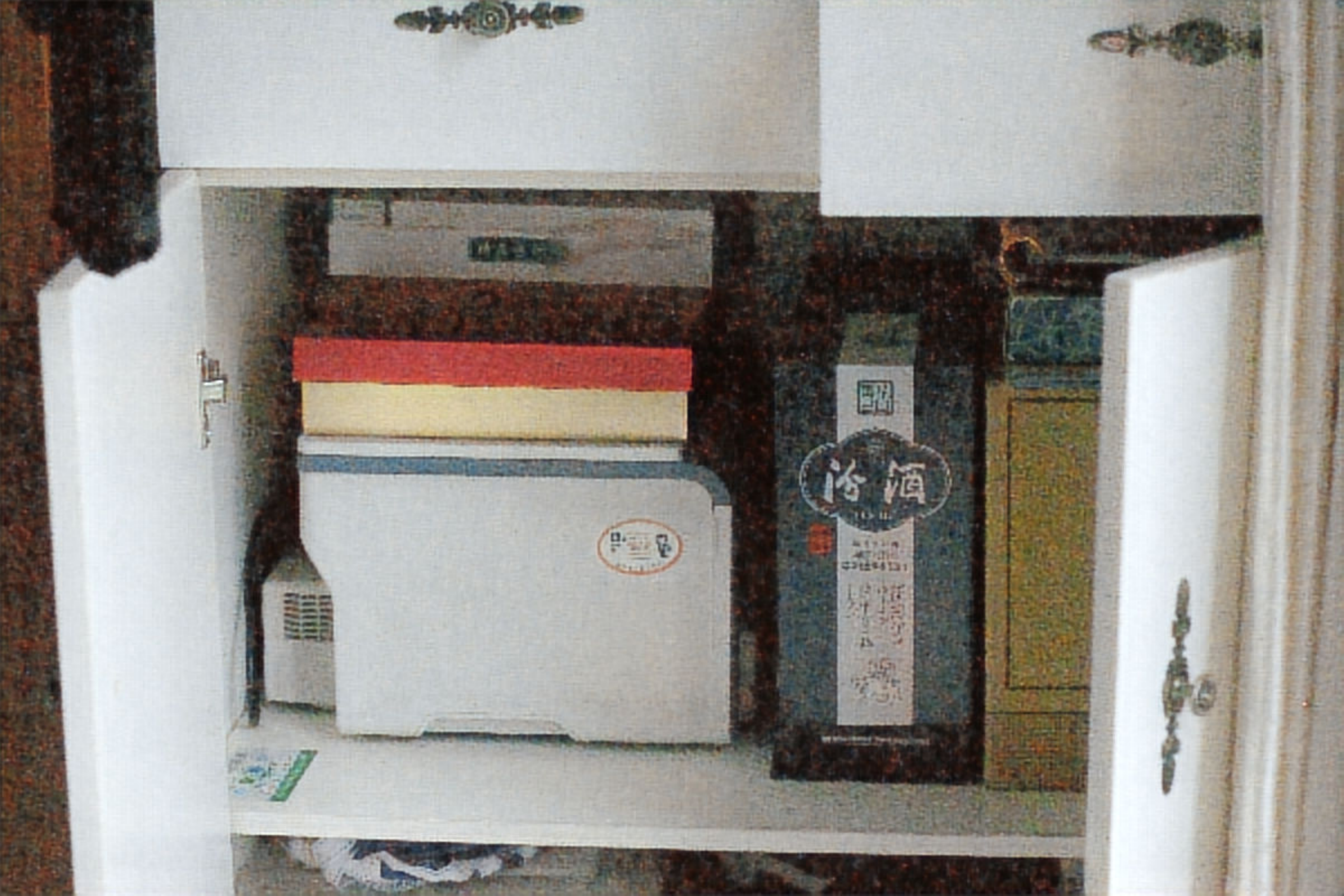}&
			\includegraphics[width=0.24\linewidth]{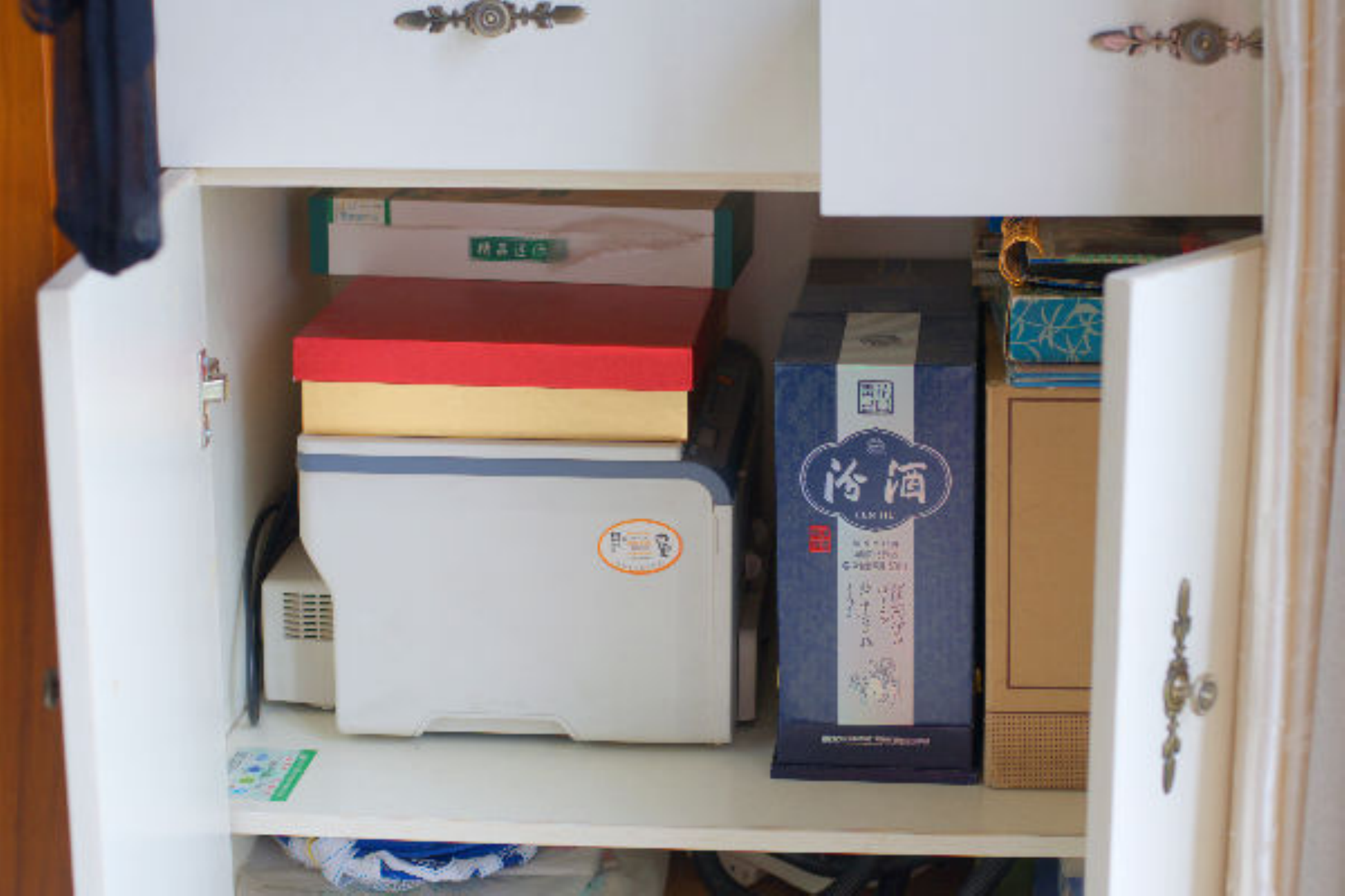}\\
			(e) GLAD\cite{GLAD} & (f) KinD\cite{KinD}  & (g) Ours & (h) Ground Truth\\
		\end{tabular}
	\end{center}
	\vspace{-0.5cm}
	\caption{Visual comparison with state-of-the-art methods on the LoL dataset.}
	\label{fig:comparelol}
	\vspace{-0.2cm}
\end{figure*}

\vspace{5pt}
\noindent\textbf{Detail Refinement}

As discussed above, relying on the upsampling module in LE alone is not enough for recovering high-resolution and detail-rich images from low-resolution inputs since it is incapable of bringing any additional information. To tackle this problem, we propose the Detail Refinement (DR) branch. DR can reconstruct fine-grained detail information under the low-resolution input, which also helps the learning of LE. Specifically, feature maps extracted from DR are transformed into feature vectors and applied to guide the learning of LE. DR and LE share the same feature extractor, as shown in Figure \ref{fig:architecture}. We propose an assumption that although low-light images are barely visible, there are still small differences among pixels. Based on this assumption, we apply super-resolution on low-light images to extract detail information, and these details will also be brightened by LE. As a result, we supervise the training of LE and DR by sharing an encoder and comparing to their respective ground truth images. The encoder is forced to learn to extract both tasks' feature.

\vspace{5pt}
\noindent\textbf{Feature Fusing module}

As aforementioned, DR contains more detail information than LE. Therefore, we design a Feature Fusing module for aggregating the features extracted from two branches and utilize fine-grained features from DR to guide the learning of LE. Precisely, the Feature Fusing module adaptively assigns different weights to different channels guided by fine-grained DR features. The weighted sum of these features is calculated to obtain the fused features before sent to the decoder of LE for brightening.

\subsection{Optimization}
\label{sec:Optimization}

The whole objective function consists of a Huber loss for Light Enhancement, a Mean Squared Error (MSE) loss for Detail Refinement, and a Color loss to relieve color distortion.

\vspace{5pt}
\noindent\textbf{Huber Loss} 

The Huber loss function describes the penalty incurred by an estimation procedure. Huber (1964) defines the loss function piecewise by \cite{huberloss}:
\begin{equation}
L_{\delta}(a) = \begin{cases}
	\frac12 a^2, & for\ |a|\leq\delta \\
	\delta |a|- \frac12\delta^2, & otherwise.
\end{cases}
\end{equation}

The variable $a$ often refers to the residuals, that is to the difference between the ground truth and output of LE $a=I_{GT}-I_{LE}$, so the former can be expanded to
\begin{footnotesize}
\begin{equation}
L_{Huber} = \begin{cases}
	\frac12 (I_{GT}-I_{LE})^2, & for\ |I_{GT}-I_{LE}|\leq\delta \\
	\delta |I_{GT}-I_{LE}|- \frac12\delta^2, & otherwise.
\end{cases}
\end{equation}
\end{footnotesize}

Two commonly used loss functions are the Mean Squared Error (MSE) loss and the Mean Absolute Error (MAE) loss. The MSE loss has the disadvantage that it tends to be dominated by outliers. One substantial problem with MAE loss is its large constant gradient, leading to missing minima at the end of training.

As defined above, the Huber loss function is convex in a uniform neighborhood of its minimum $a=0$; at the boundary of this uniform neighborhood, the Huber loss function has a differentiable extension to an affine function at points $a=-\delta$ and $a=\delta$. These properties allow it to combine the sensitivity of MSE loss and the robustness of MAE loss.

\begin{figure*}[h]
	\begin{center}
		\begin{tabular}{c@{ }c@{ }c@{ }c@{ }c@{ }c}
			\includegraphics[width=.19\textwidth]{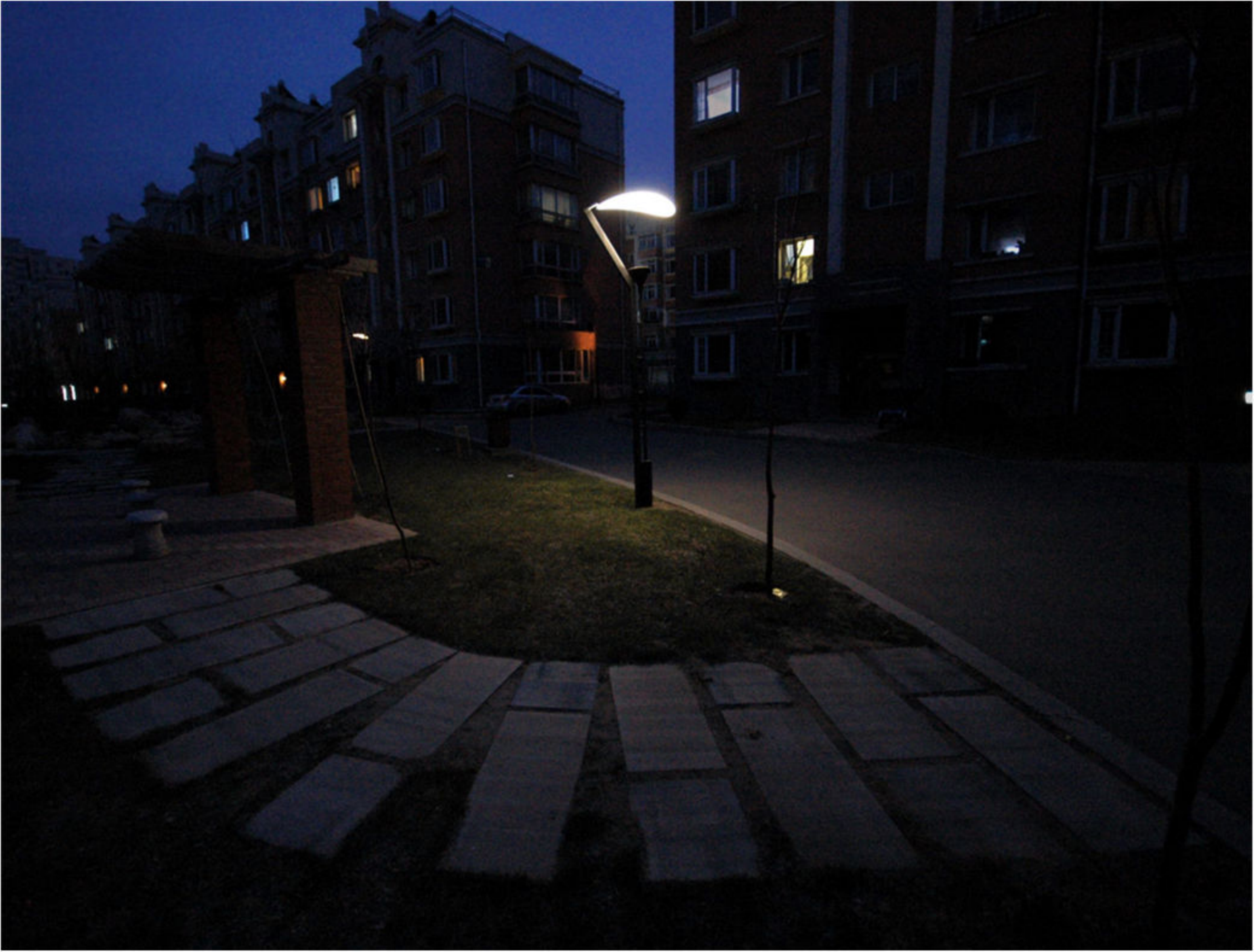}~&
			\includegraphics[width=.19\textwidth]{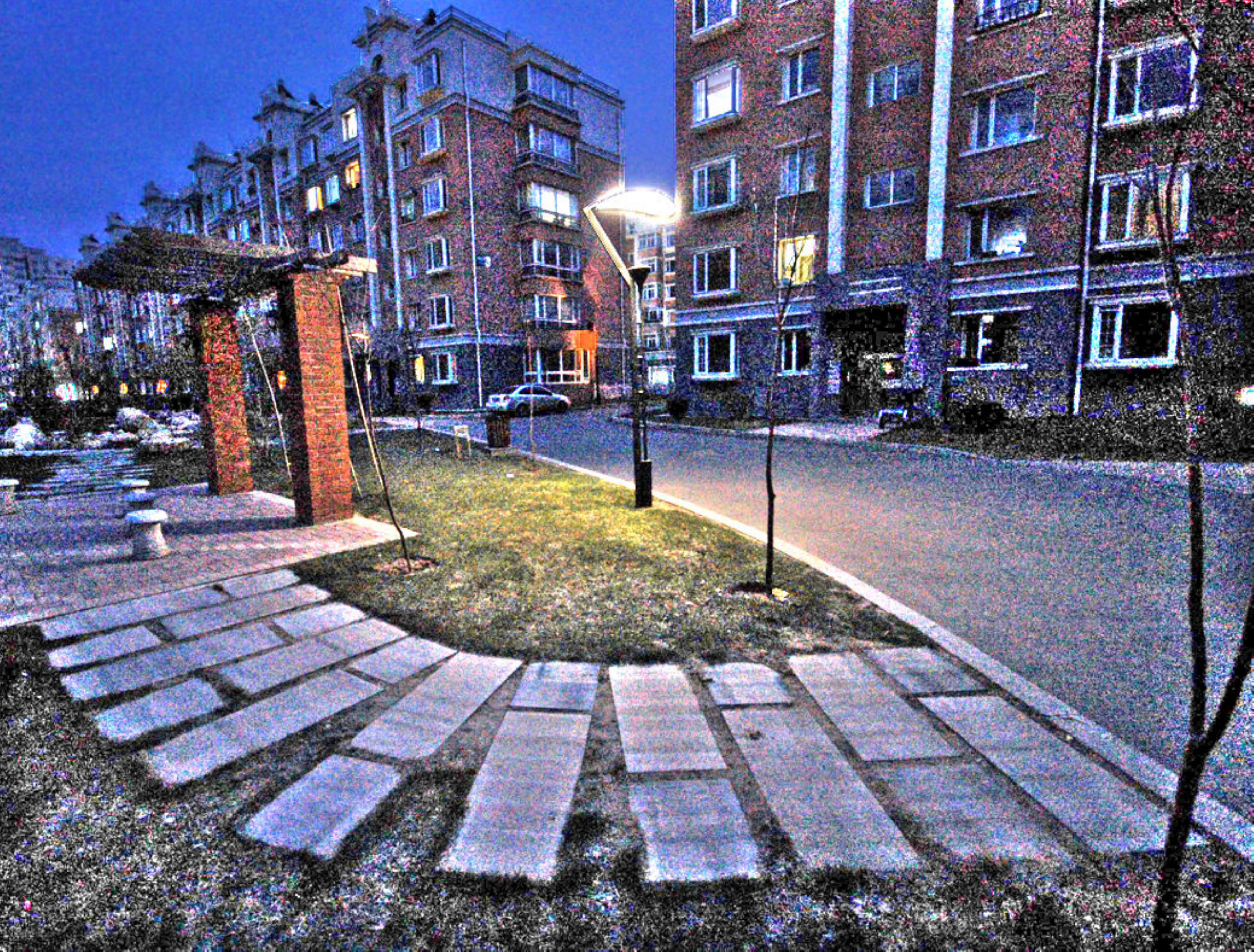}~&
			\includegraphics[width=.19\textwidth]{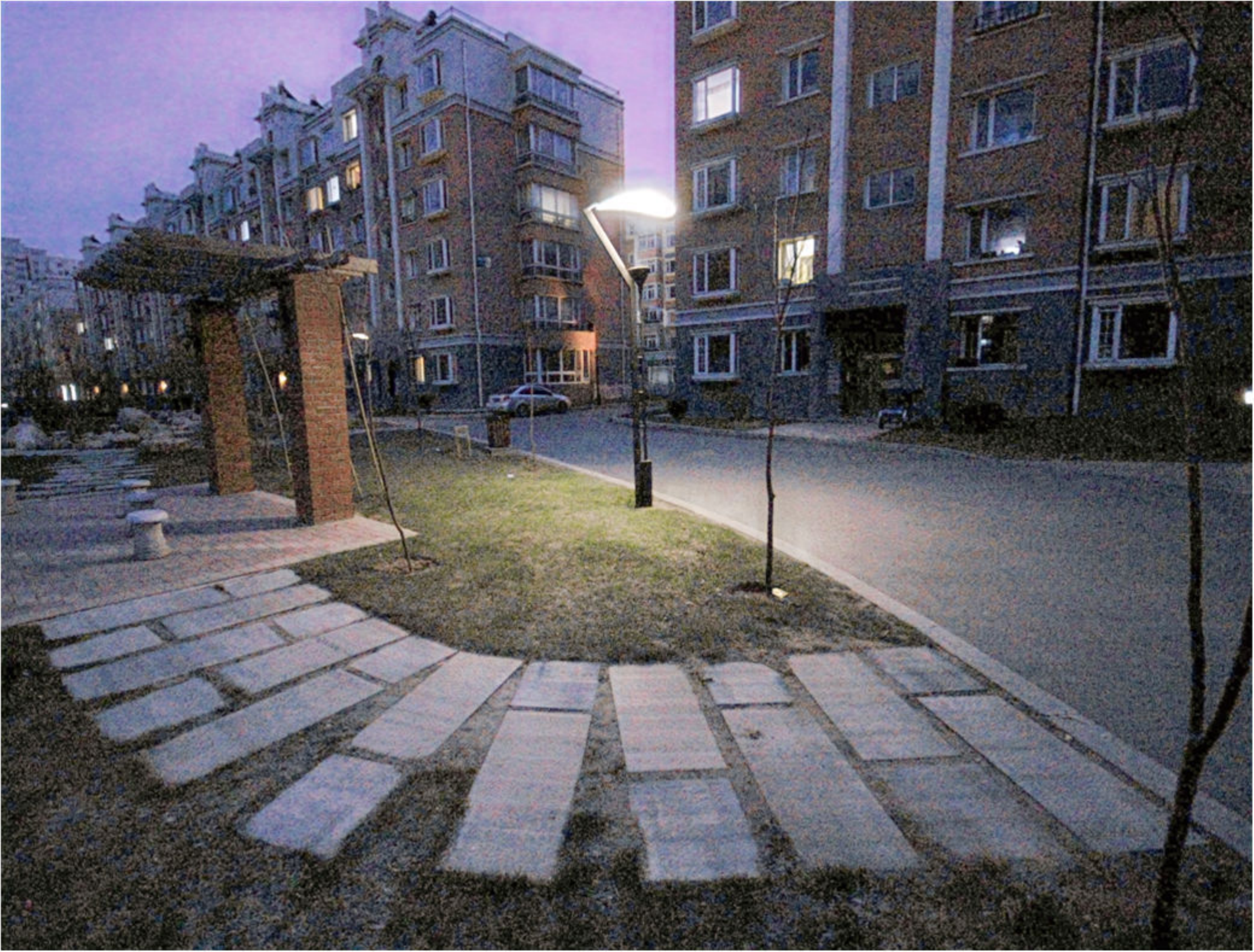}~&
			\includegraphics[width=.19\textwidth]{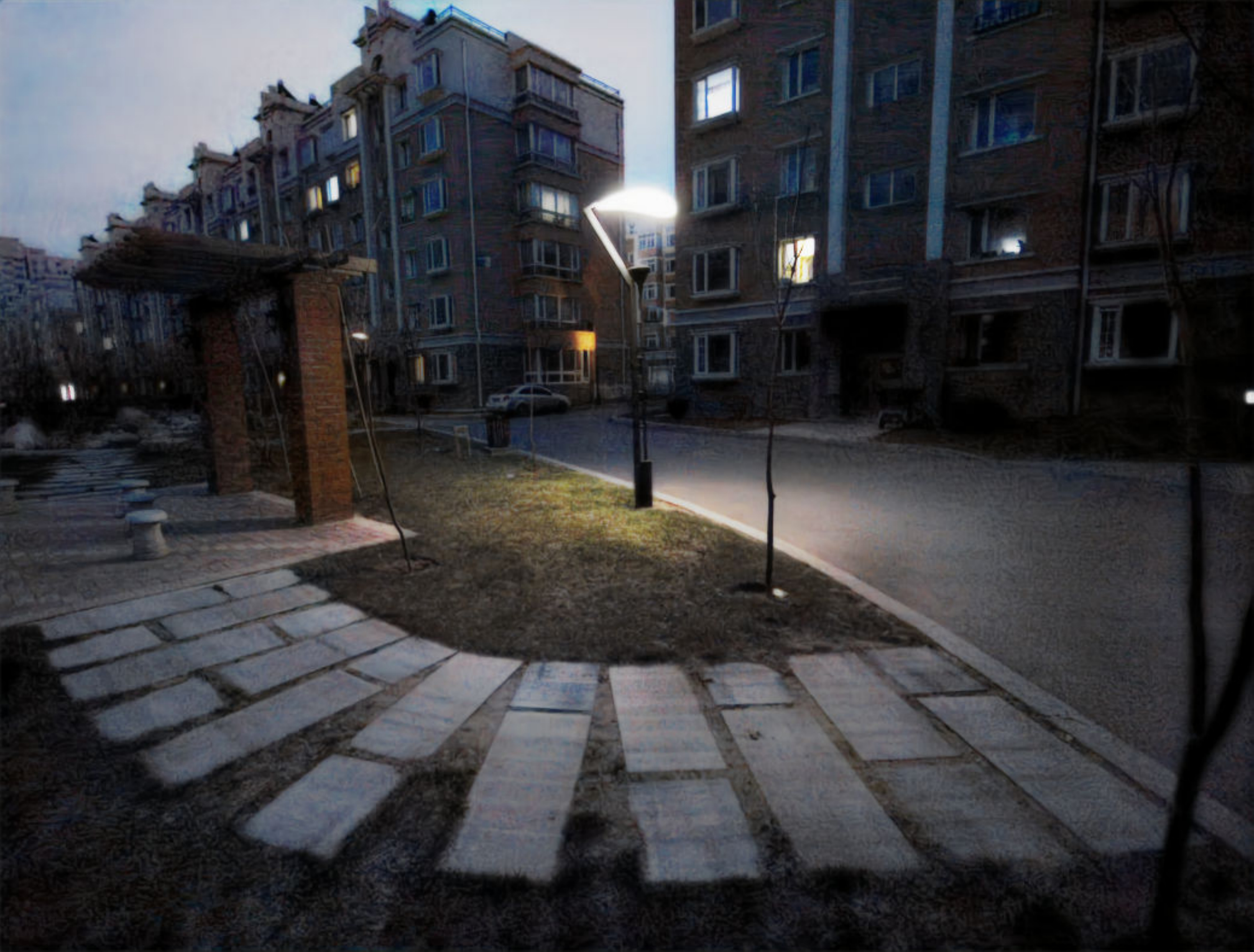}~&
			\includegraphics[width=.19\textwidth]{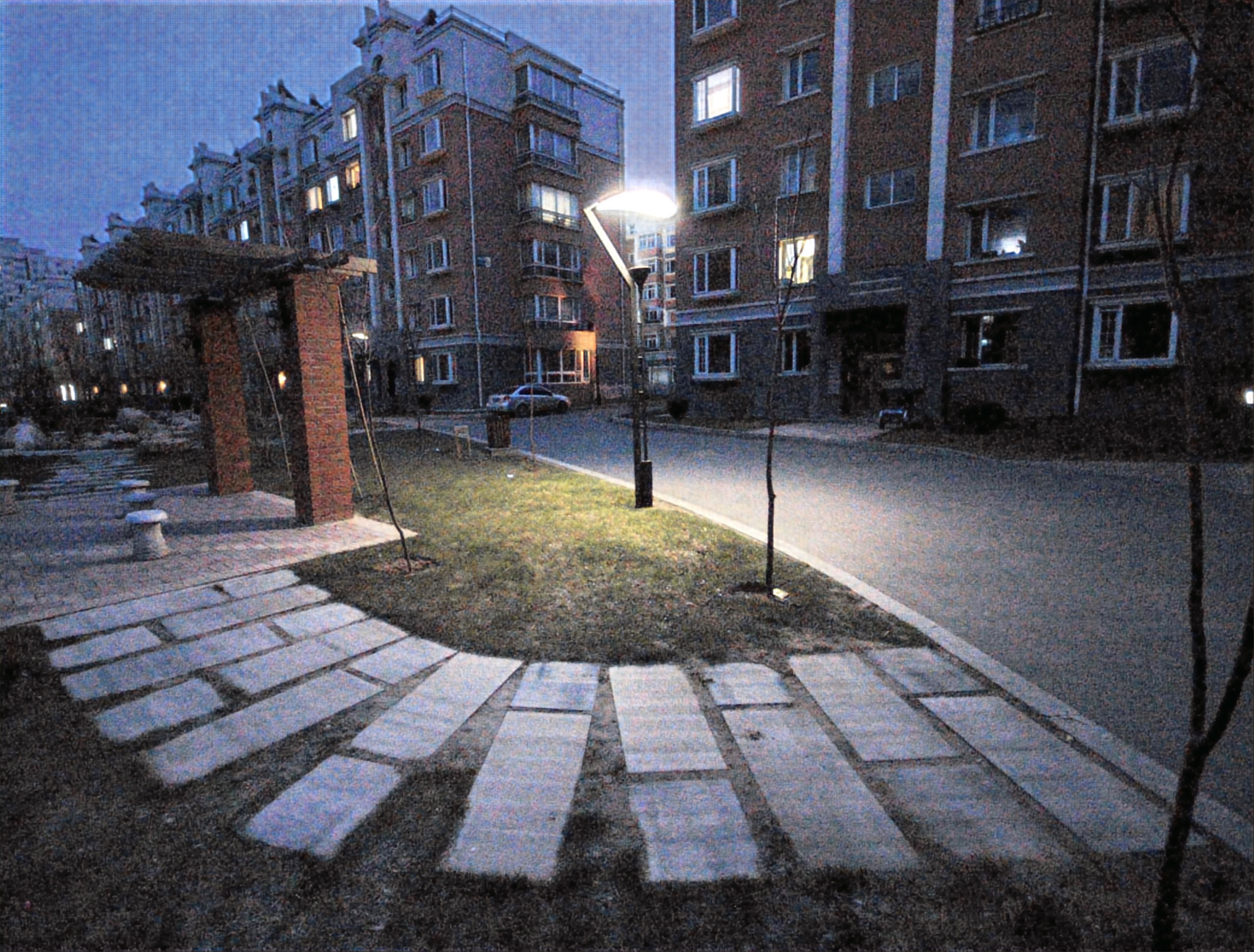}\\
			(a) Input~& (b) LIME~\cite{LIME}& (c) GLAD~\cite{GLAD}& (d) KinD~\cite{KinD}& (e) Ours\\
		\end{tabular}
	\end{center}
	\vspace{-0.5cm}
	\caption{Visual comparison with state-of-the-art methods on the LIME dataset.}
	\vspace{-0.2cm}
	\label{fig:comparelime}
\end{figure*}

\begin{figure*}[h]
	\begin{center}
		\begin{tabular}{c@{ }c@{ }c@{ }c@{ }c@{ }c}
			\includegraphics[width=.19\textwidth]{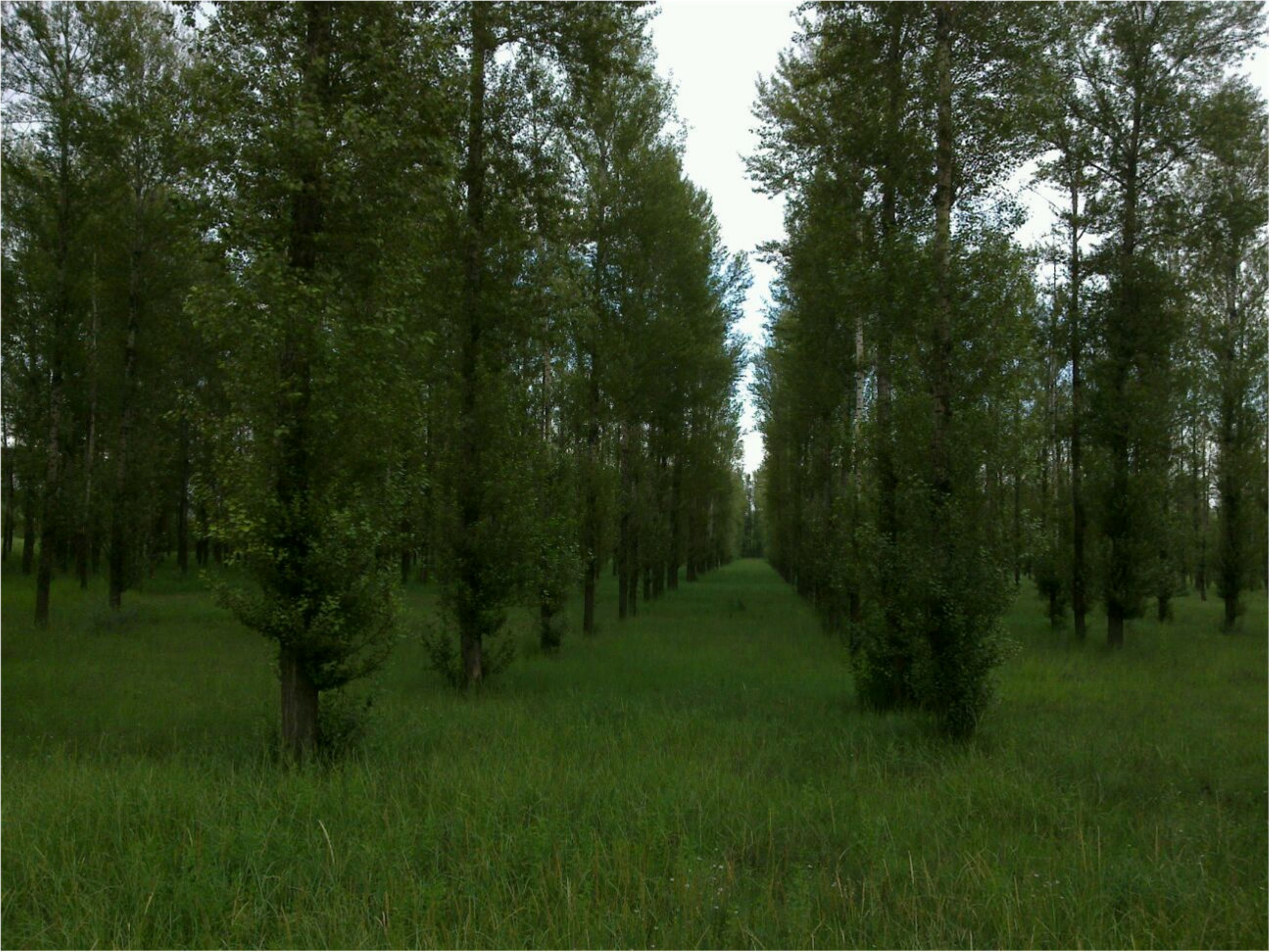}~&
			\includegraphics[width=.19\textwidth]{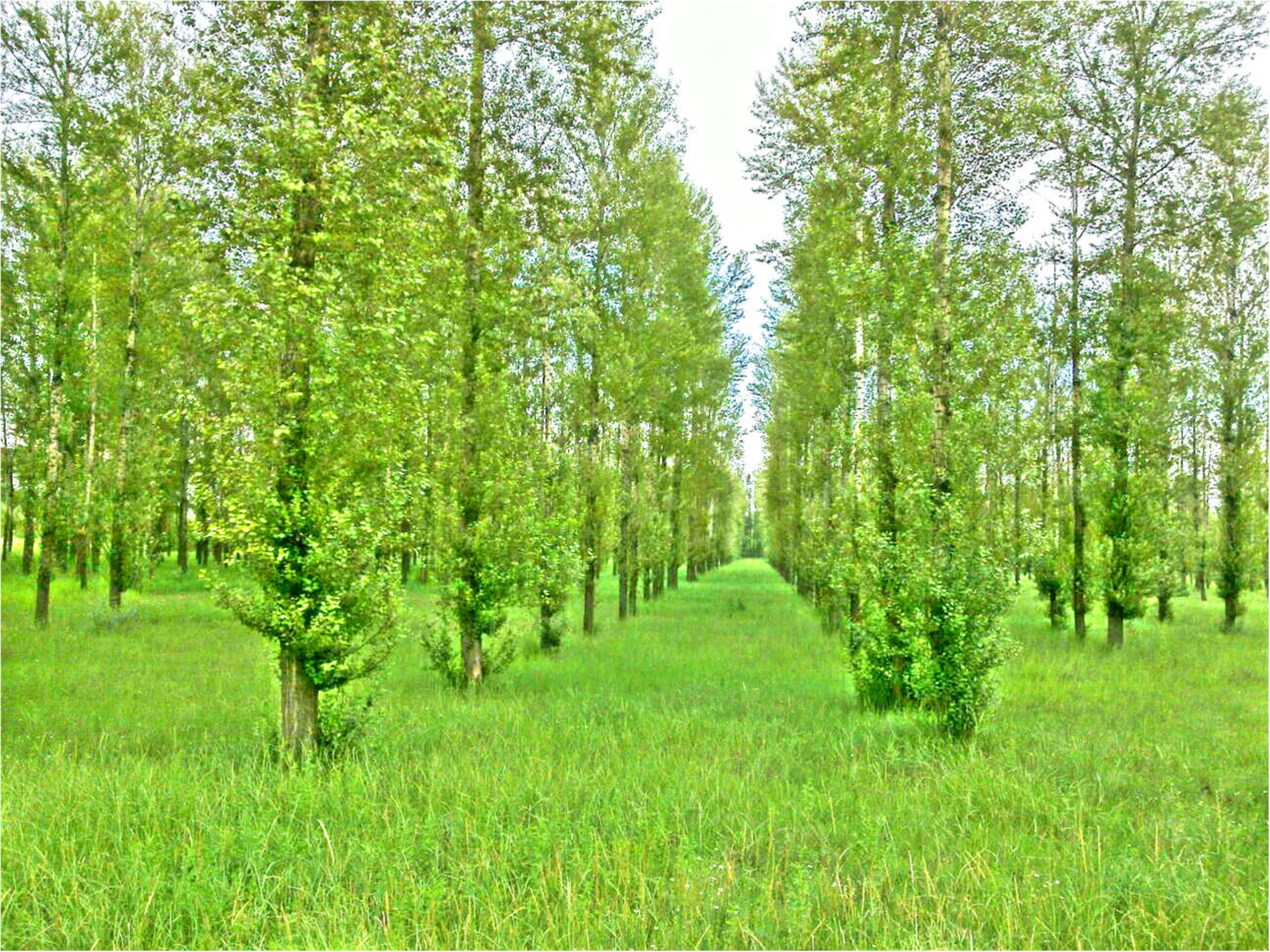}~&
			\includegraphics[width=.19\textwidth]{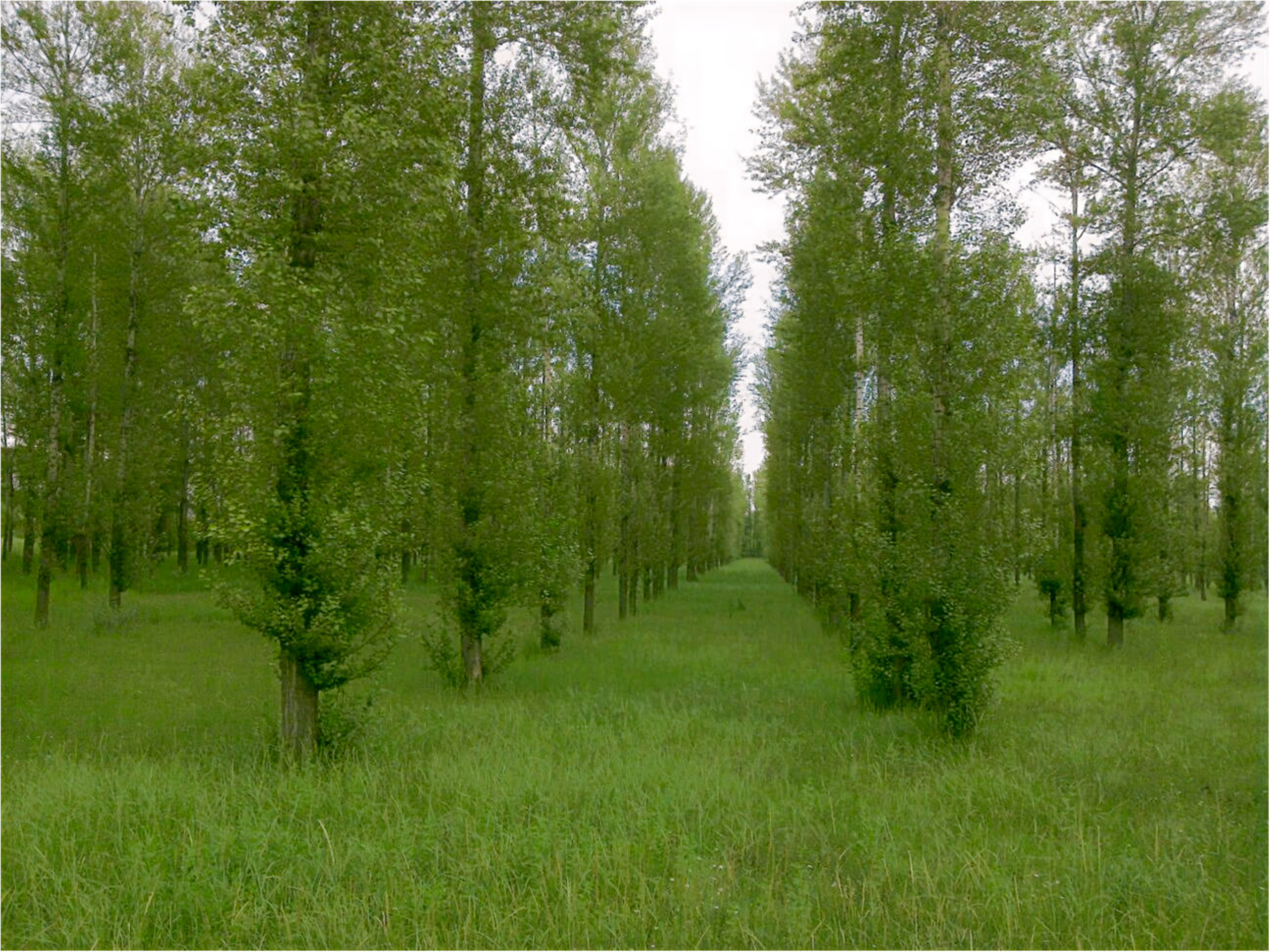}~&
			\includegraphics[width=.19\textwidth]{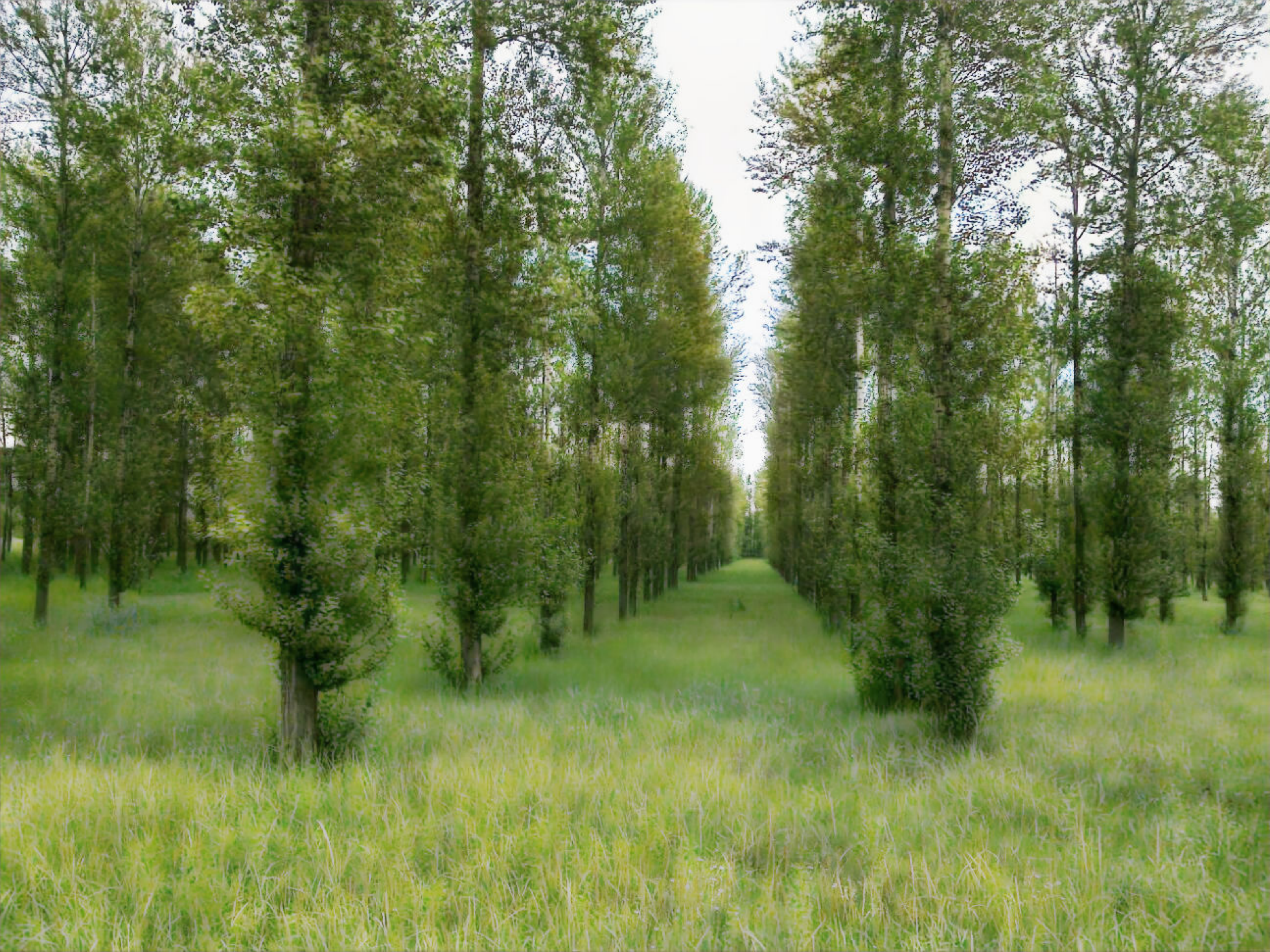}~&
			\includegraphics[width=.19\textwidth]{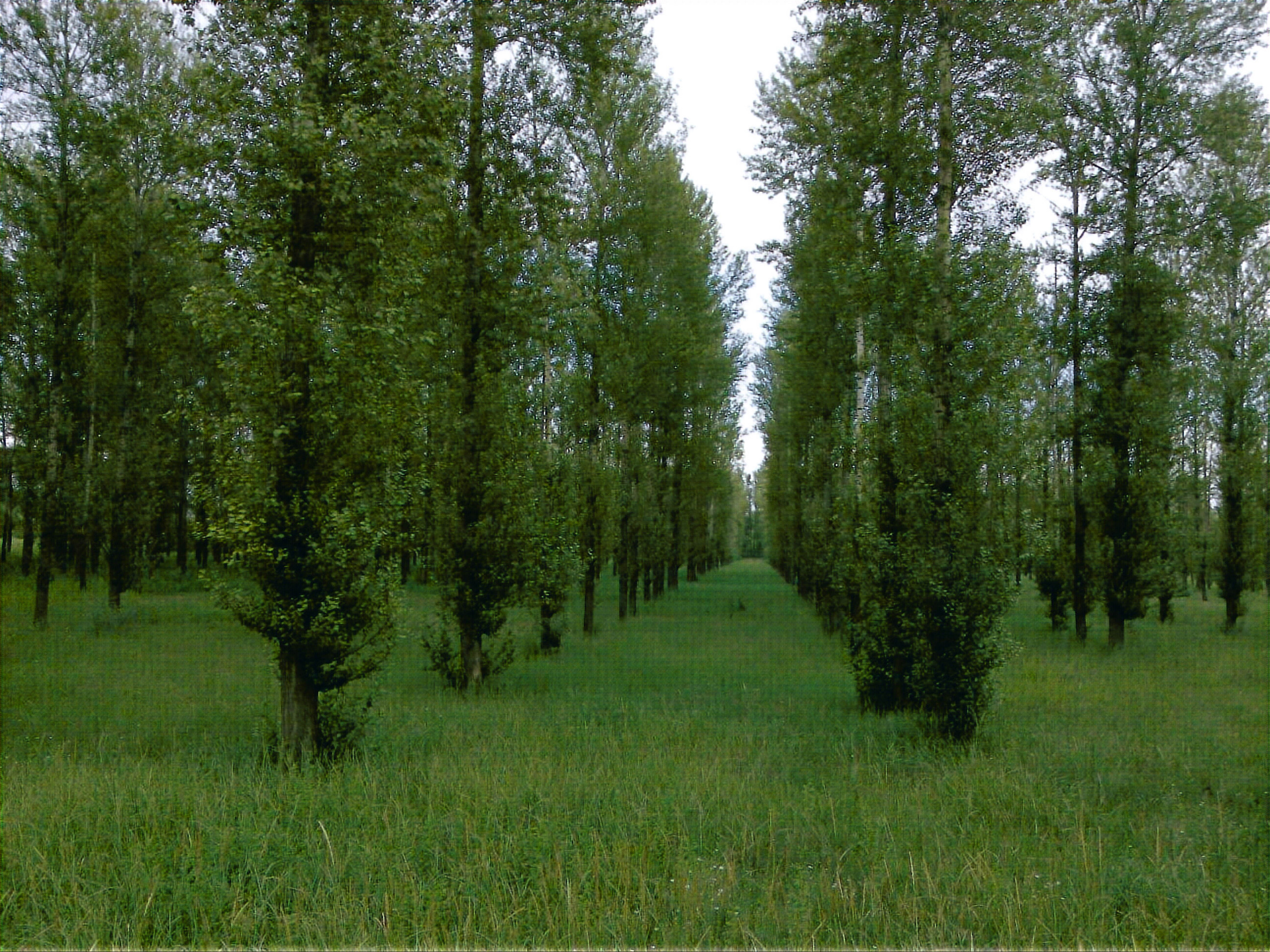}\\
			(a) Input~& (b) LIME~\cite{LIME}& (c) GLAD~\cite{GLAD}& (d) KinD~\cite{KinD}& (e) Ours\\
		\end{tabular}
	\end{center}
	\vspace{-0.5cm}
	\caption{Visual comparison with state-of-the-art methods on the NPE dataset.}
	\vspace{-0.2cm}
	\label{fig:comparenpe}
\end{figure*}

\vspace{5pt}
\noindent\textbf{MSE Loss}

For DR, we apply a conventional MSE loss to supervise its training.
\begin{equation}
L_{MSE} = \frac1N \sum_{i=1}^{n}||I_{DR}-I_{GT}||^2
\end{equation}

\begin{figure*}[h]
	\begin{center}
		\begin{tabular}{c@{ }c@{ }c@{ }c@{ }c@{ }c}
			\includegraphics[width=.19\textwidth]{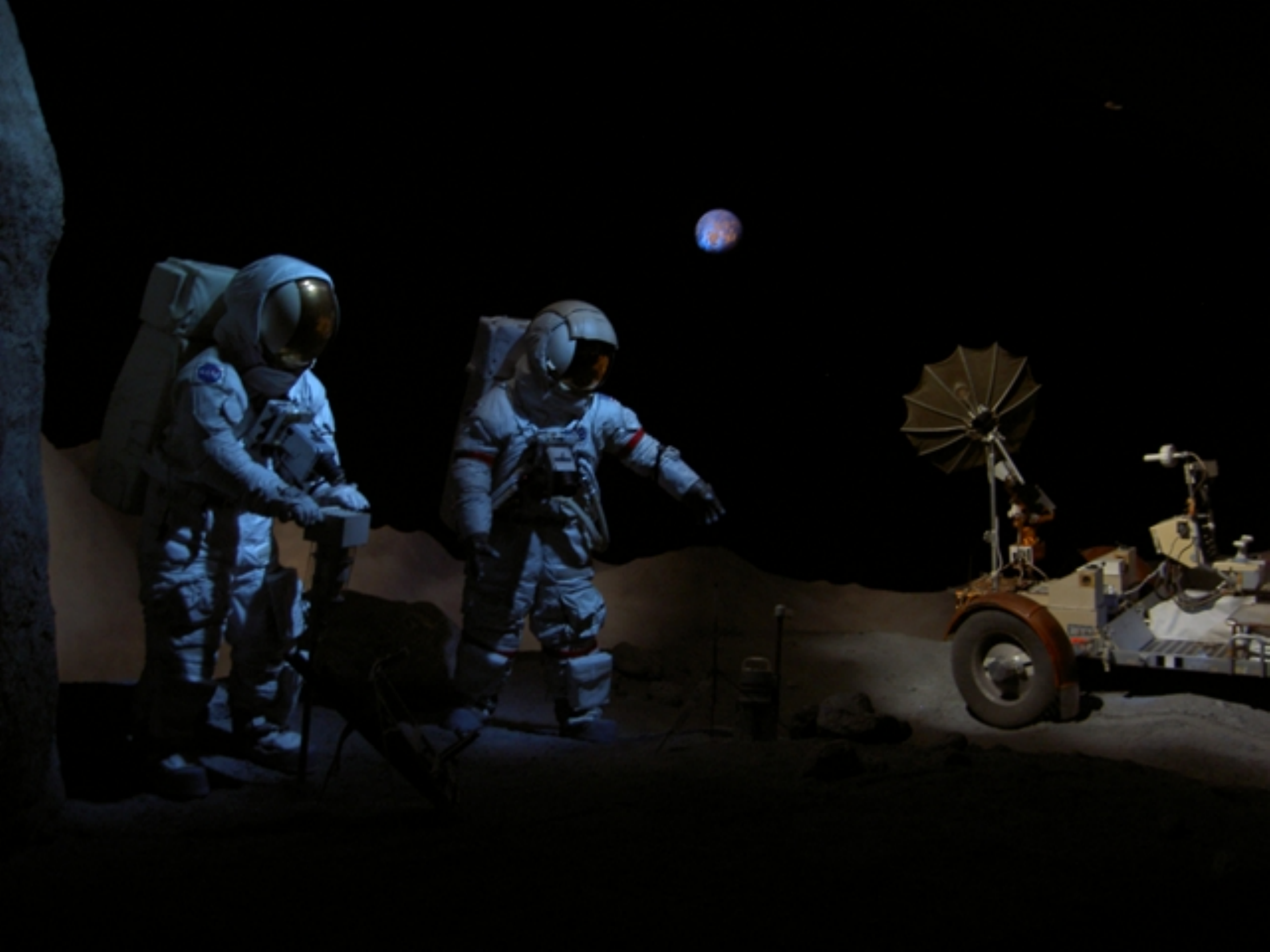}~&
			\includegraphics[width=.19\textwidth]{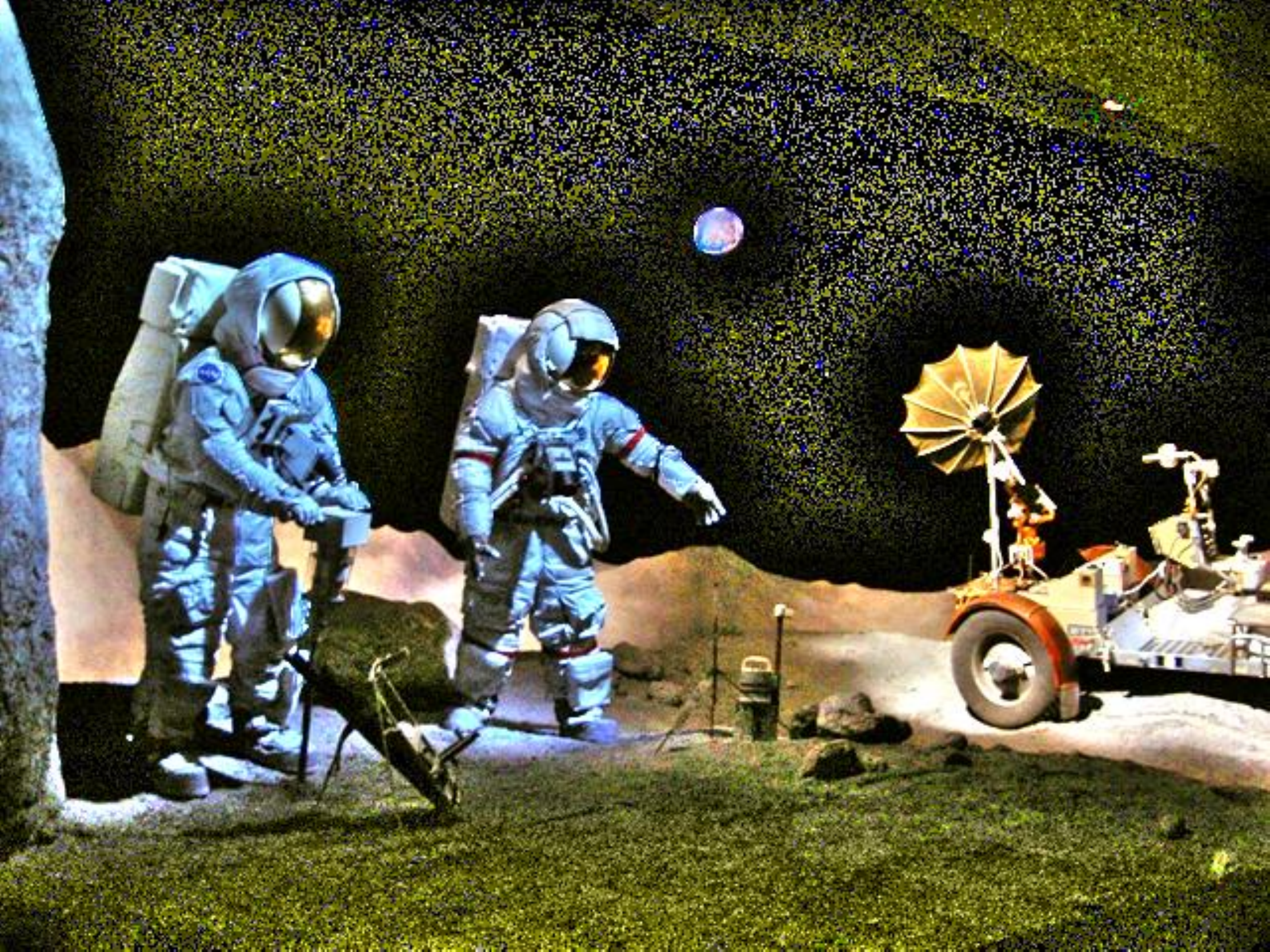}~&
			\includegraphics[width=.19\textwidth]{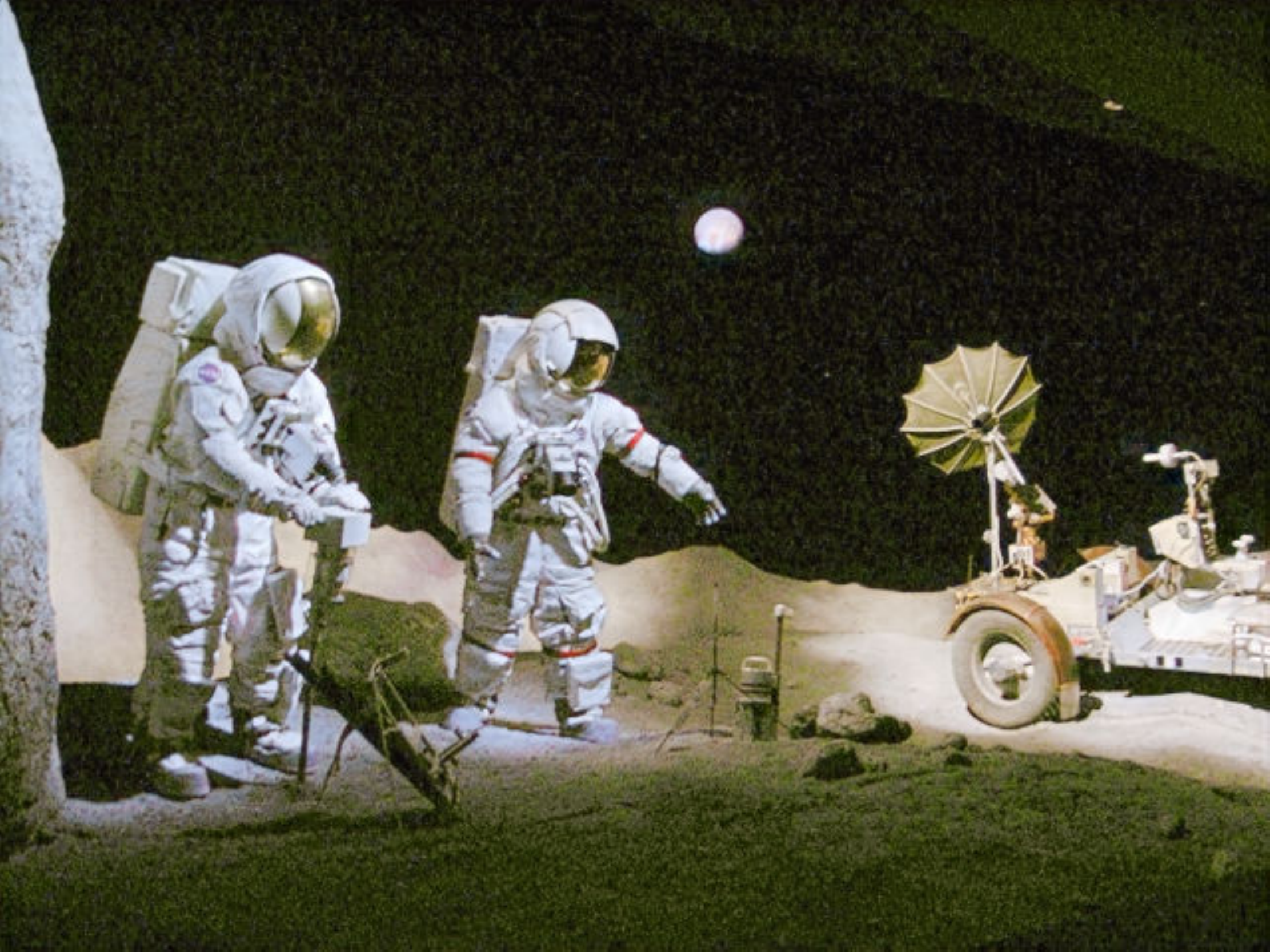}~&
			\includegraphics[width=.19\textwidth]{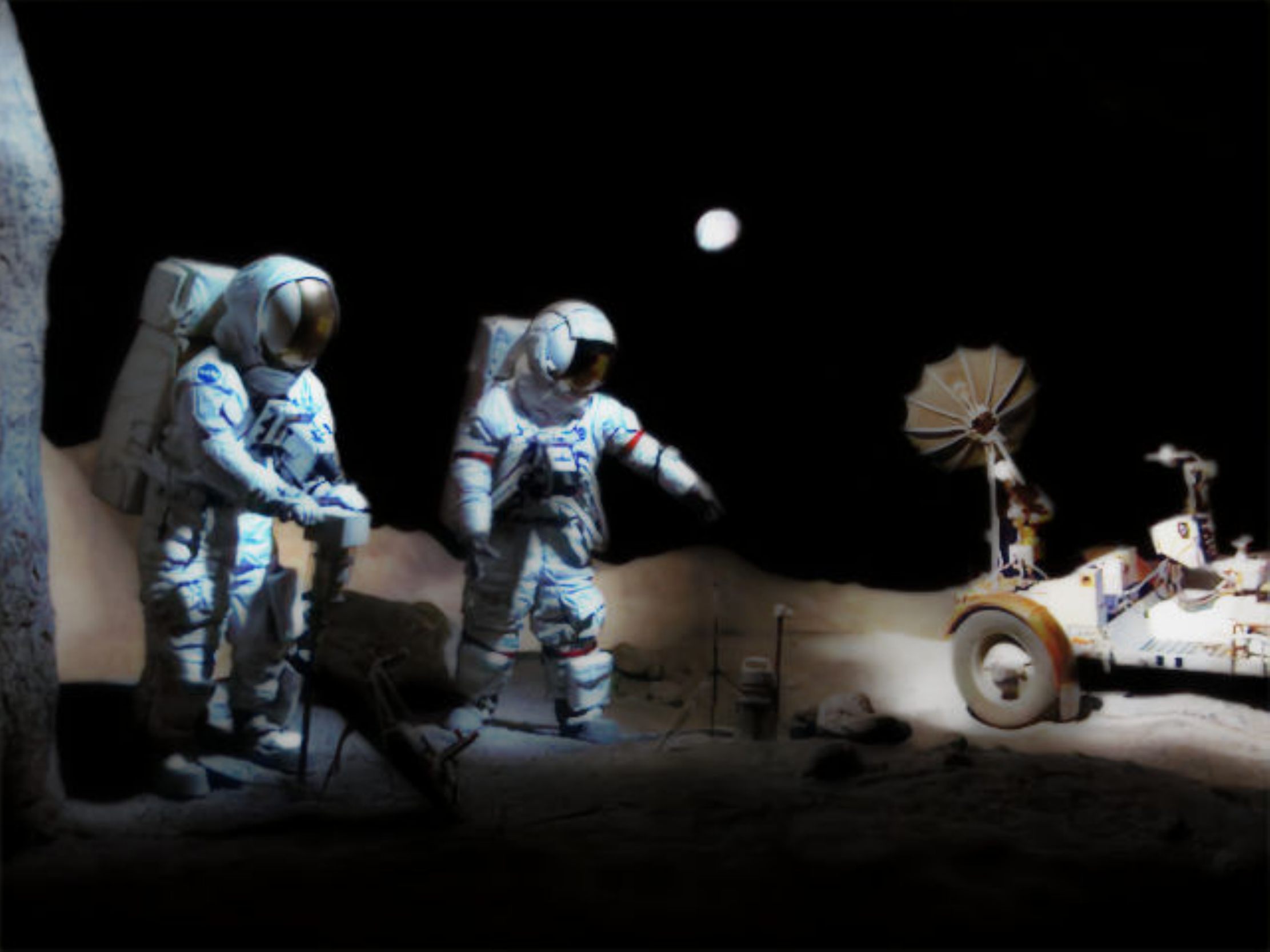}~&
			\includegraphics[width=.19\textwidth]{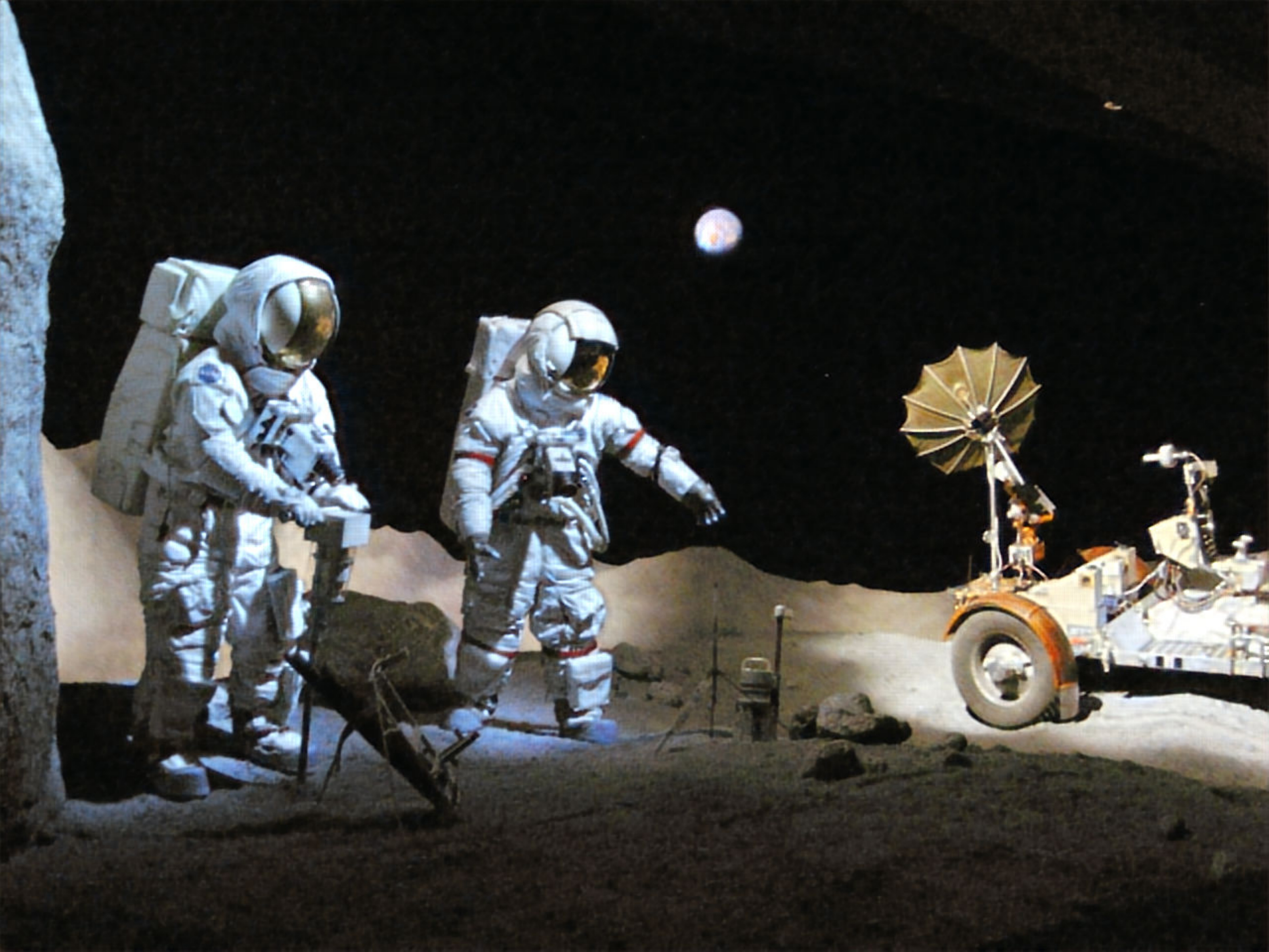}\\
			(a) Input~& (b) LIME~\cite{LIME}& (c) GLAD~\cite{GLAD}& (d) KinD~\cite{KinD}& (e) Ours\\
		\end{tabular}
	\end{center}
	\vspace{-0.5cm}
	\caption{Visual comparison with state-of-the-art methods on the DICM dataset.}
	\vspace{-0.2cm}
	\label{fig:comparedicm}
\end{figure*}

\vspace{5pt}
\noindent\textbf{Color Loss}

It has been noticed that existing low-light image enhancement methods suffer from color distortion. We minimize a Color loss in HSV color space, calculating the cosine distance between the predicted image and ground truth image in Hue and Saturation channel. The loss function is expressed as:
\begin{equation}
L_{Color} = L_H + L_S
\end{equation}

\begin{equation}
L_H = 1-cos(H_p, H_{GT})
\end{equation}

\begin{equation}
L_S = 1-cos(S_p, S_{GT})
\end{equation}
where $cos(,)$ is an operation to calculate the cosine similarity. $H_p$ and $H_{GT}$ refer to the Hue channel of the prediction and ground truth, respectively. Similarly, $S_p$ and $S_{GT}$ refer to the Saturation channel.

\begin{figure*}[h]
	\vspace{0.6cm}
	\begin{center}
		\begin{tabular}{c@{ }c@{ }c@{ }c@{ }c@{ }}
			\includegraphics[width=0.3\linewidth]{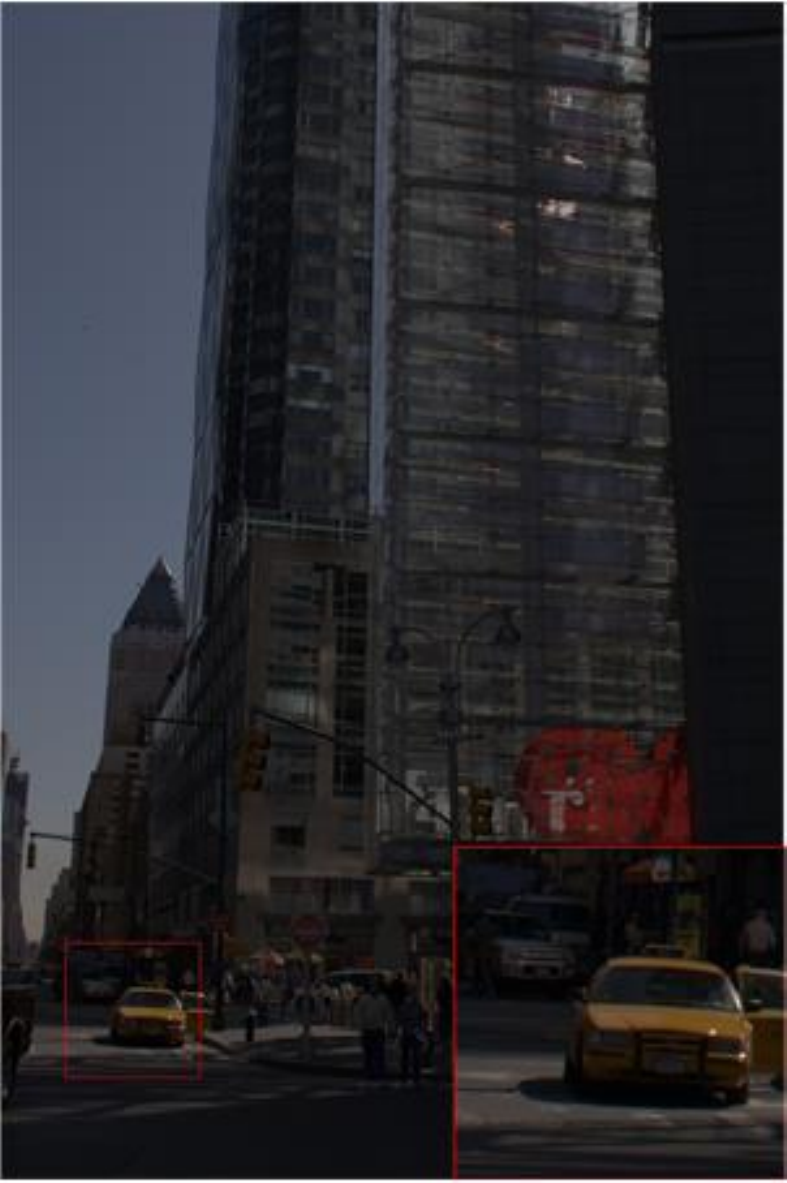}&
			\includegraphics[width=0.3\linewidth]{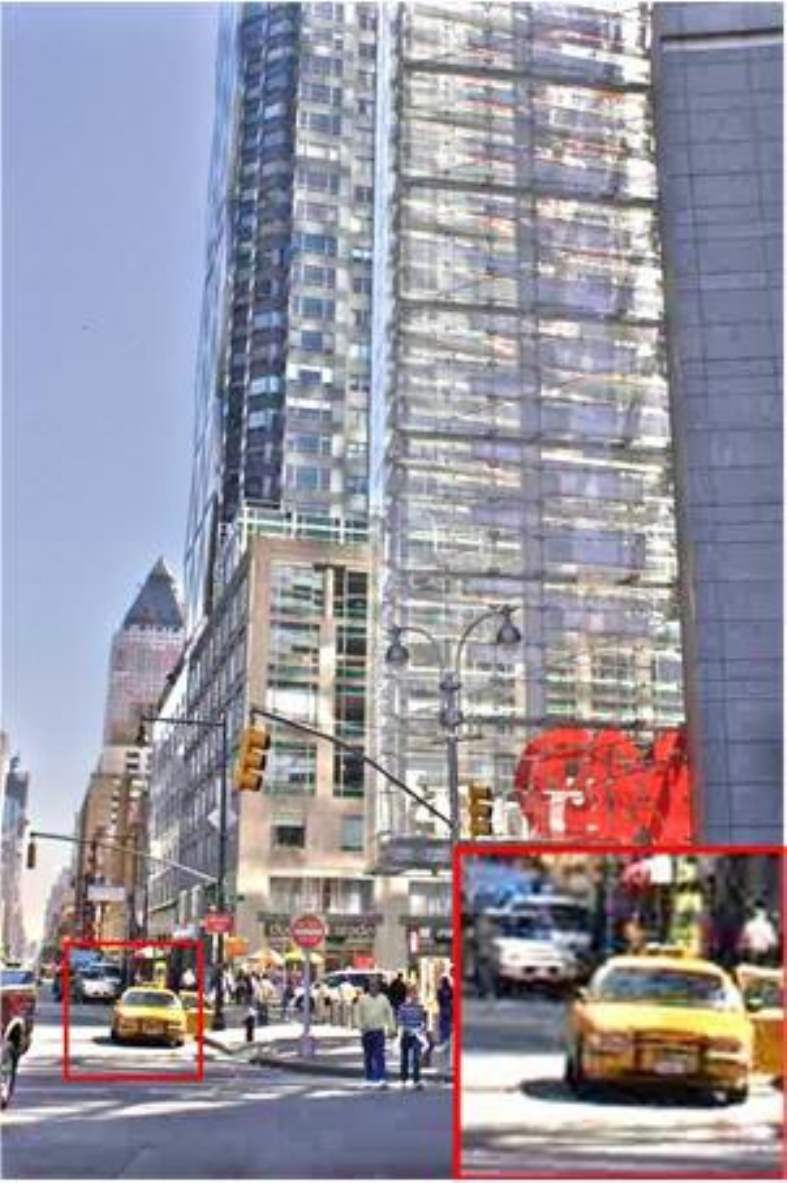}&
			\includegraphics[width=0.3\linewidth]{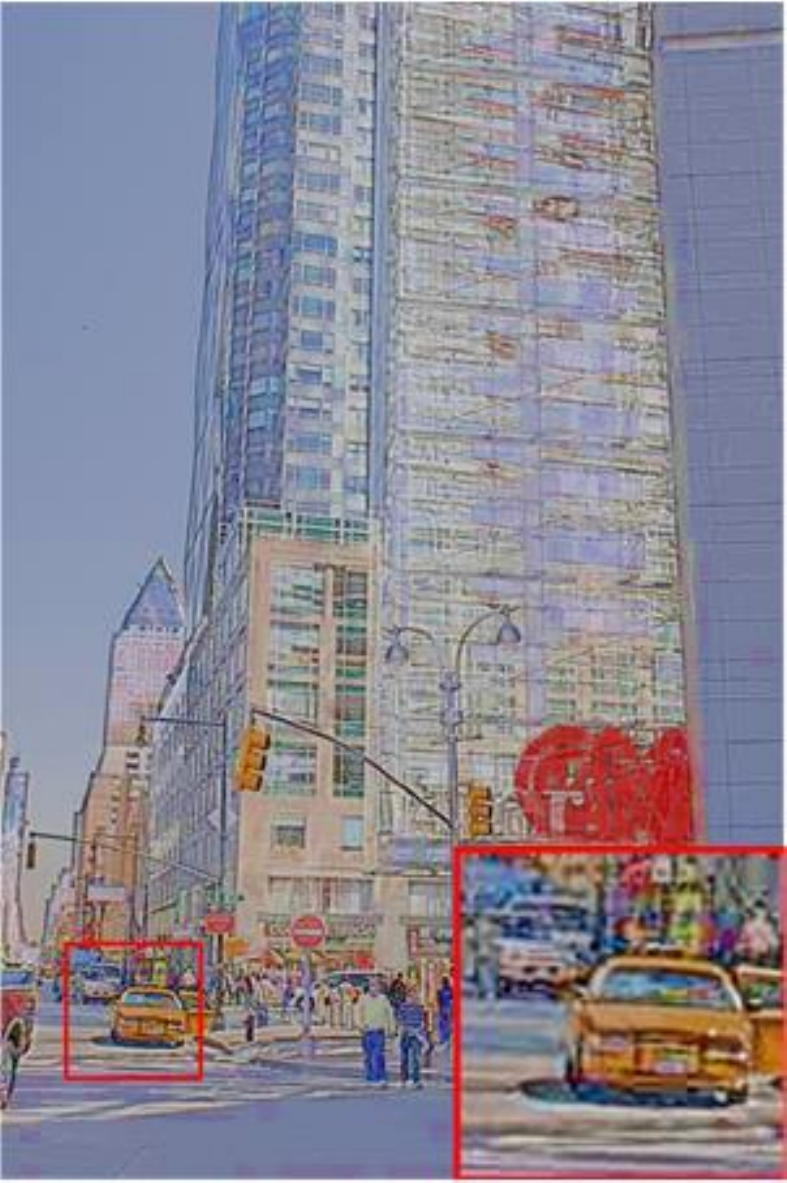}\\
			(a) Input & (b) LIME\cite{LIME} & (c) Retinex-Net\cite{Retinex-Net} \\
			\includegraphics[width=0.3\linewidth]{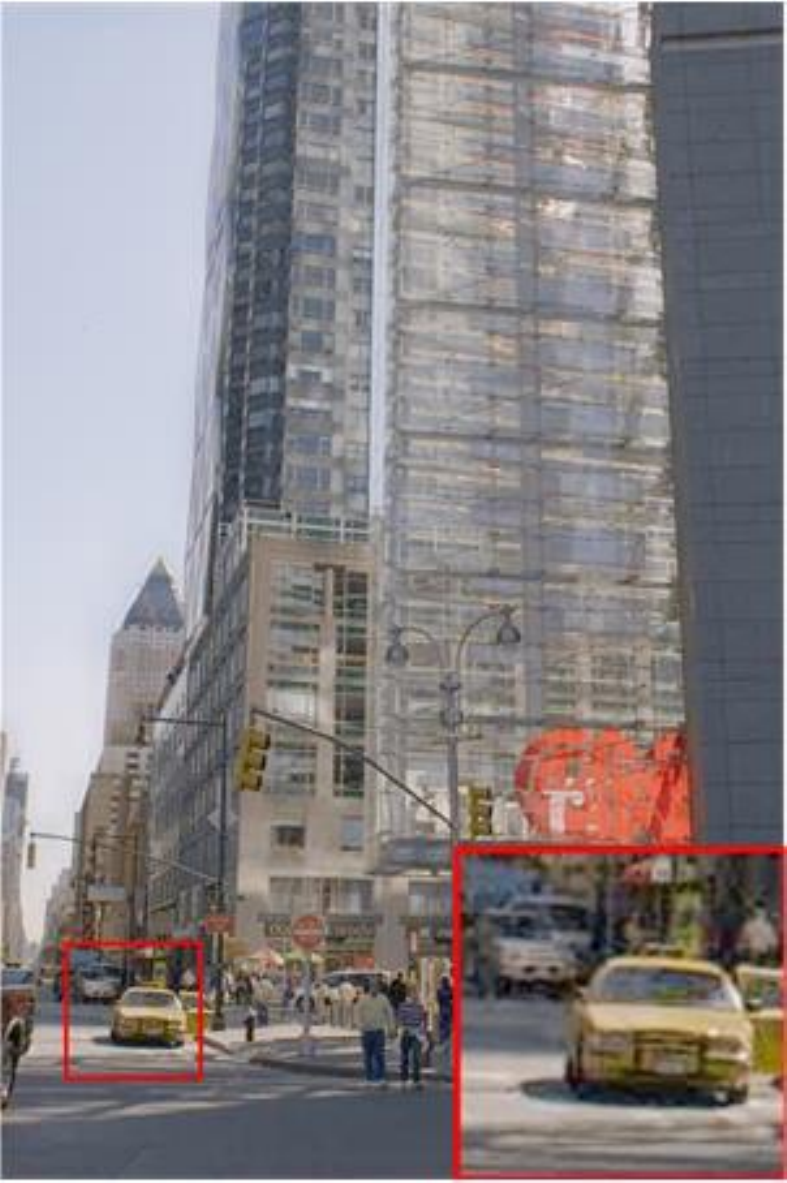}&
			\includegraphics[width=0.3\linewidth]{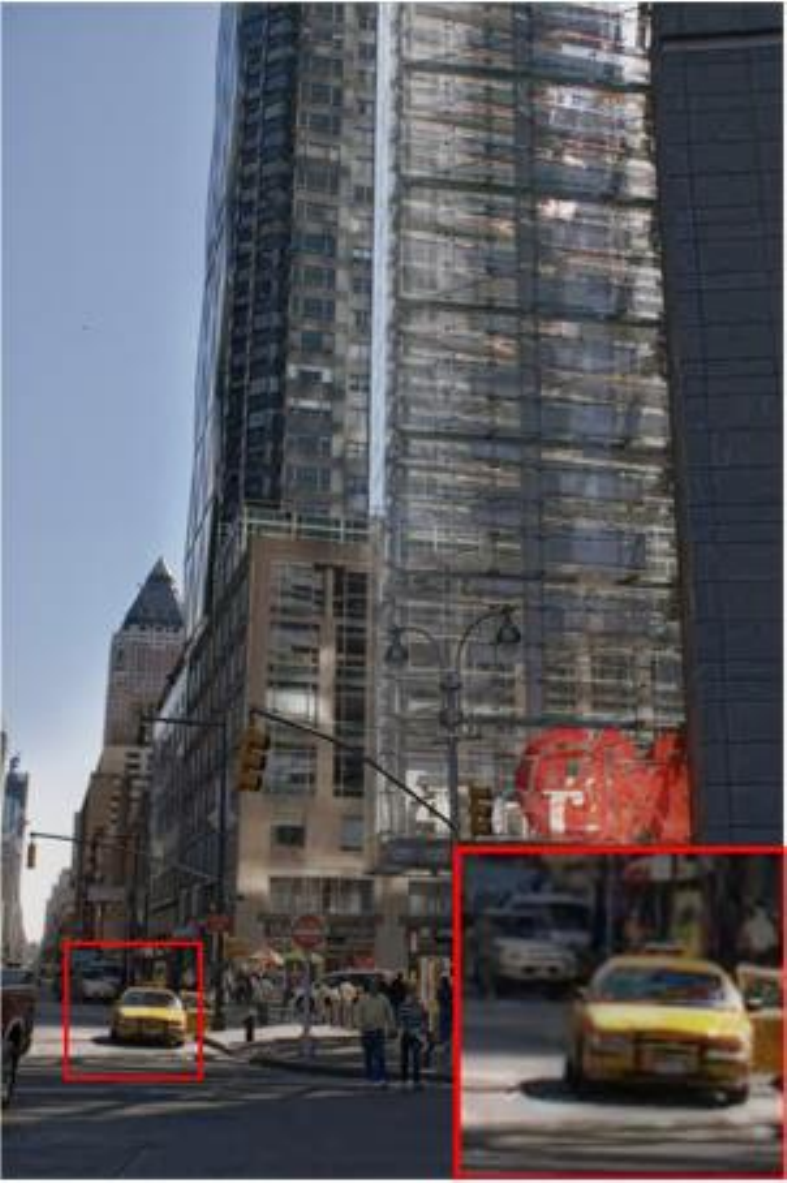}&
			\includegraphics[width=0.3\linewidth]{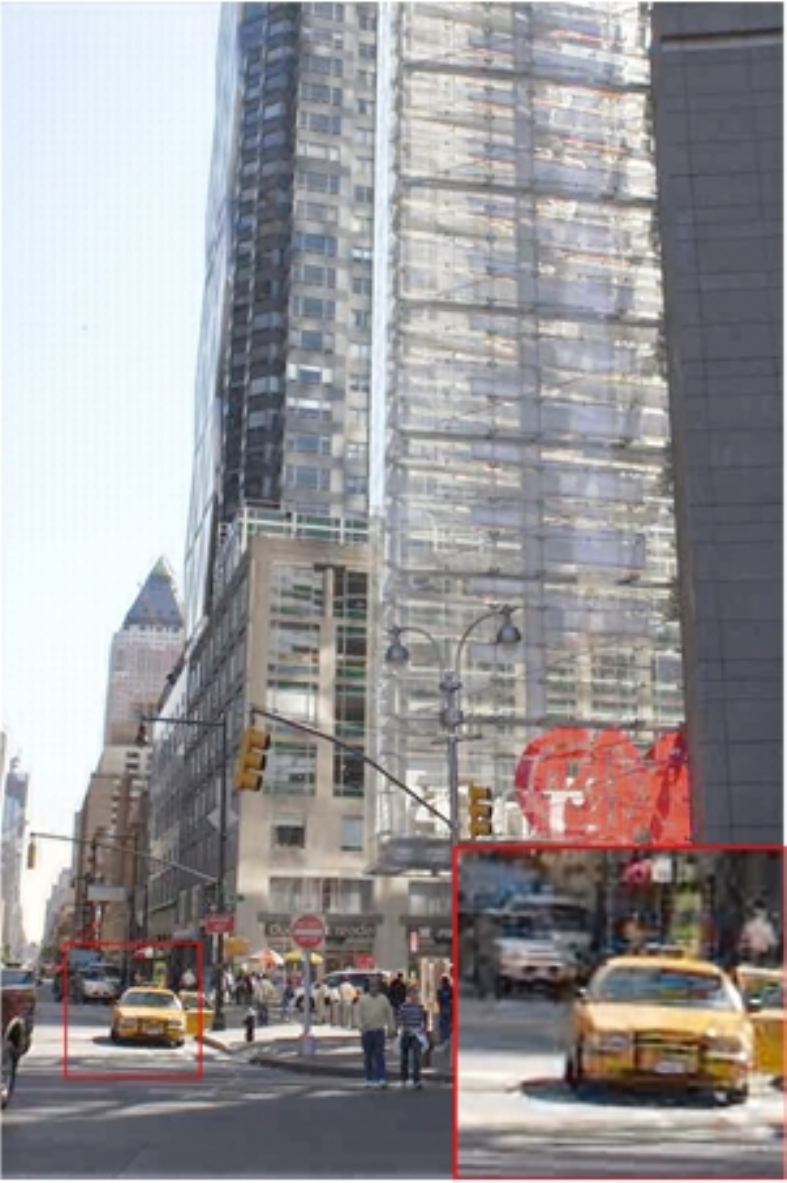}&\\
			(d) GLAD\cite{GLAD} & (e) KinD\cite{KinD}  & (f) Ours \\
		\end{tabular}
	\end{center}
	\vspace{-0.5cm}
	\caption{Detail comparison with state-of-the-art methods on the MIT Adobe 5K dataset.}
	\label{fig:compare5k}
\end{figure*}

\vspace{5pt}
\noindent\textbf{Total loss}

The total loss is the combination of all three losses above:
\begin{equation}
\label{con:totalloss}
L_{total} = L_{Huber} + \lambda_1 L_{MSE} + \lambda_2 L_{Color}
\end{equation}
where $\lambda_i$ is the weight of corresponding loss. We minimize the total loss end-to-end.

\section{Experiments}

In this part, we perform qualitative and quantitative assessments of the results and compare them with state-of-the-art methods. Firstly, we describe the datasets and provide the implementation details. Then, results for low-light image enhancement are reported.

\vspace{3pt}
\subsection{Datasets and Evaluation Metrics}

We use the LoL\cite{lol} dataset as the training dataset, which includes 500 image pairs. Each image pair consists of a low-light input image and its corresponding well-exposed reference image. We also assess the proposed algorithm’s ability to restore details on the MIT-Adobe 5K dataset\cite{mit5k}. It contains 5000 images captured with DSLR cameras. The tonal attributes of all images are manually adjusted by five trained photographers (labeled as experts A to E). Same as in \cite{hu2018exposure,park2018distort,wang2019underexposed}, we also consider the enhanced images of expert C as the ground-truth. Moreover, we use some no-reference datasets to evaluate the qualitative performance, including LIME\cite{LIME}, DICM\cite{DICM} and NPE\cite{NPE}.

We employ two common metrics (i.e., PSNR and SSIM) to quantitatively evaluate the color and structural similarity between the predicted results and the corresponding ground truth images. High PSNR and SSIM values correspond to reasonably good results. 

\vspace{3pt}
\subsection{Implementation Details} 

The proposed model is end-to-end trained with the Adam optimizer for 2000 epochs. The initial learning rate is set to $10^{-4}$. It first decreases to $10^{-5}$ at 500-th epoch and then to $10^{-6}$ at 1000-th epoch. A batch size of 8 is applied. For data augmentation, we randomly cropped patches of size 256$\times$256 from normal-light images followed by vertical flips and rotations by multiples of 90 degrees. For the model input, we downsample the ground truth patches to 128$\times$128 with bicubic interpolation. 

The filter weights of each layer are initialized with a standard zero mean and 0.02 standard deviation Gaussian function. Bias is initialized as a constant. Both weights of the MSE loss and the Color loss are set to 0.1 to balance the scale of losses. The entire network is trained on an NVidia GTX 2080Ti GPU and Intel Core i7-8700 3.20GHz CPU using the Tensorflow framework.

\begin{table}[ht]
	\begin{center}
		\caption{Results of the ablation study.}
		\begin{tabular}{c|c}
			\toprule[1.5pt]
			& \\[-8pt]
			\textbf{Architecture} & \textbf{PSNR} \\ \midrule[1pt]
			& \\[-8pt]
			U-net architecture & 19.60 \\
			& \\[-8pt]
			LE & 20.78 \\ 
			& \\[-8pt]
			LE + DR & 21.73 \\ 
			& \\[-8pt]
			LE + DR + FF & \textbf{22.83} \\  \bottomrule[1.5pt]
		\end{tabular}
		\label{tab:ablation}
	\end{center}
	\vspace{-0.8cm}
\end{table}

\vspace{3pt}
\subsection{Comparison with state-of-the-art methods}

This section demonstrates the effectiveness of our algorithm by evaluating it for the low-light image enhancement task. We report PSNR/SSIM values of our method and several other techniques in Table \ref{tab:compare_lol} for the LoL\cite{lol} dataset. It can be seen that our NEID achieves significant improvements over previous approaches. Notably, when compared to the recent best methods, NEID obtains about 2 dB performance gain over KinD\cite{KinD} on the LoL dataset.

Moreover, we evaluate the visual effect on widely-adopted datasets, including LoL\cite{lol}, LIME\cite{LIME}, NPE\cite{NPE} and DICM\cite{DICM}. Some of the above methods are involved as competitors. Figure \ref{fig:comparelol}-\ref{fig:comparedicm} gives visual comparisons of images with different contents. From the results, although these methods can brighten the inputs, severe visual degradations caused by the obstinate blur and color distortion remain. Compared to these techniques, our method generates more natural and vivid images.

It is inappropriate to measure the proposed model’s ability to restore details with the above datasets since they include few detail-rich images. Therefore, we choose MIT Adobe 5K, a high-resolution dataset with rich details in images. Figure \ref{fig:compare5k} shows the comparison of details between resluts of our NEID and other methods. We can observe that, compared with other methods, NEID successfully restores images with rich and distinct details, which manifests the effectiveness of our method.

It is worth mentioning that the input images are first downsampled with bicubic interpolation to generate images with the same size as other methods. In other words, the proposed NEID surpasses state-of-the-art methods with smaller size input images. As a result, our method reduces the computational costs and improves the inference speed, meeting the requirements of the scenarios described in Section \ref{sec:intro}.

\vspace{3pt}
\subsection{Ablation Study}

Besides quantitatively and qualitatively evaluate our method, we also assess the effectiveness of different components in the proposed NEID. As shown in Table \ref{tab:ablation}, taking the U-net architecture as the baseline, the LE branch can improve the performance from 19.60 dB to 20.78 dB. By adding the DR branch, the PSNR can be effectively improved by about 1 dB. While combining with the Feature Fusing module, the performance can be further improved to 22.83 dB (3.23 dB higher than the baseline), indicating that transferring the detail information from DR to LE is of pivotal importance. The ablation study demonstrates the necessity of components in the NEID network.

\section{Conclusion}

In this work, we have proposed a practical framework named NEID for low-light enhancement. Inspired by super-resolution, the proposed network increases the resolution of output normal-light images to meet the requirements of field-to-command scenarios. The LE branch enhances the illumination and details of input images, with the resolution of output images increased. The DR branch uses the same encoder as the LE branch and refines detail information. The Feature Fusing (FF) module is introduced to fuse the features of two branches, guiding the learning of LE with detail features extracted from DR. Extensive experiments demonstrated the distinct advantages of our design over the state-of-the-art alternatives.

\noindent\textbf{Acknowledgements.}
This work was supported in part by National Key Research and Development Program under Grant 2019YFC1511404, in part by National Natural Science Funding(No.62002026), and in part by MoE-CMCC ``Artificial Intelligence'' Project under Grant MCM20190701.

{\small
\bibliographystyle{ieee_fullname}
\bibliography{liuc}
}

\end{document}